\documentclass[twocolumn,nopacs,preprintnumbers,superscriptaddress,amsmath,amssymb,jcp]{revtex4}

\usepackage[pdftitle={appendix}, pdfauthor={Tobias Schmidt}, bookmarks=true]{hyperref}
\usepackage{graphicx}% Include figure files
\usepackage{epstopdf}
\usepackage{dcolumn}% Align table columns on decimal point
\usepackage{bm}% bold math
\usepackage{color}
\usepackage[usenames,dvipsnames]{xcolor}
\usepackage{paralist}
\usepackage{verbatim}
\usepackage[normalem]{ulem}
\newcommand{\sandw}[3]{\langle#1|#2|#3\rangle}

\newcommand{\sgn}[0]{\textrm{sgn}}
\newcommand{\eps}[0]{\varepsilon}
\newcommand{\pphi}[0]{\varphi}
\newcommand{\up}[0]{\uparrow}
\newcommand{\dn}[0]{\downarrow}
\newcommand{\lp}[0]{\left}
\newcommand{\rp}[0]{\right}
\newcommand{\s}[0]{\sigma}
\newcommand{\de}[0]{\partial}
\newcommand{\KS}[0]{{K\!S}}

\newcommand{\half}[0]{\frac{1}{2}}
\newcommand{\rarr}[0]{\rightarrow}
\newcommand{\rr}[0]{(\mathbf{r})}

\newcommand{\darsec}[0]{\texttt{DARSEC}}

\newcommand{\hartree}[0]{Ha}

\newcommand{\rb}[0]{\mathbf{r}}
\newcommand{\red}[1]{\textcolor{black}{#1}}
\newcommand{\redd}[1]{\textcolor{black}{#1}}

\begin{document}

\title{Effect of ensemble generalization on the highest-occupied Kohn-Sham eigenvalue}

\author{Eli Kraisler}
\thanks{These authors contributed equally}
\affiliation{Department of Materials and Interfaces, Weizmann Institute of Science, Rehovoth 76100,Israel}

\author{Tobias Schmidt}
\thanks{These authors contributed equally}
\affiliation{Theoretical Physics IV, University of Bayreuth, 95440 Bayreuth, Germany}

\author{Stephan K\"{u}mmel}
\affiliation{Theoretical Physics IV, University of Bayreuth, 95440 Bayreuth, Germany}

\author{Leeor Kronik}
\affiliation{Department of Materials and Interfaces, Weizmann Institute of Science, Rehovoth 76100,Israel}

\date{\today}% It is always \today, today, but any date may be explicitly specified

\begin{abstract}

There are several approximations to the exchange-correlation functional in density-functional theory that accurately predict total energy-related properties of many-electron systems, such as binding energies, bond lengths, and crystal structures.
Other approximations are designed to describe potential-related processes, such as charge transfer and photoemission. However, the development of a functional which can serve the two purposes simultaneously is a long-standing challenge. Trying to address it, we employ in the current work the ensemble generalization procedure proposed in Phys.\ Rev.\ Lett.\ \textbf{110}, 126403 (2013). 
Focusing on the prediction of the ionization potential via the highest occupied Kohn-Sham eigenvalue, we examine a variety of  exchange-correlation approximations: the local spin-density approximation, semi-local generalized gradient approximations, and global and local hybrid functionals. Results for a test set of 26 diatomic molecules and single atoms are presented. 
We find that the aforementioned ensemble generalization systematically improves the prediction of the ionization potential, for various systems and exchange-correlation functionals, without compromising the accuracy of total energy-related properties.
We specifically examine hybrid functionals. These depend on a parameter controlling the ratio of semi-local to non-local functional components. The ionization potential obtained with ensemble-generalized functionals is found to depend only weakly on the parameter value, contrary to common experience with non-generalized hybrids, thus eliminating one aspect of the so-called `parameter dilemma' of hybrid functionals.

\end{abstract}

%\pacs{31.15.ep, 31.15.eg, 31.10.+z, 71.15.Mb}% PACS, the Physics and Astronomy Classification Scheme.
\maketitle

\section{Introduction}\label{sec.intro} %Writer: Tobi

Modern density-functional theory (DFT), based on the theoretical foundation laid by Hohenberg, Kohn, and Sham~\cite{Hohenberg1964,Kohn1965} in the 1960s, in principle provides an exact framework for treating the many-electron problem. Within this framework, the electron-electron interaction is expressed via the exchange-correlation (xc) energy term, $E_\mathrm{xc}[n]$, which is a functional of the electron density $n\rr$~\cite{Parr1989,Dreizler1995,Fiolhais2003}. In practice, one has to approximate this energy contribution, aiming at a numerically efficient yet accurate description of the electronic structure of various many-electron systems, such as atoms, molecules, and solids. 

During the eventful history of DFT, many density functional approximations (DFAs) to the exact xc energy were developed~\cite{Becke2014,Burke2012}. 
The performance of each of them can be evaluated from a two-fold perspective: On the one hand, ground-state quantities such as binding energies, bond lengths, and crystal structures are closely related to the total energy of the system and hence to the approximate xc \textit{energy} $E_\mathrm{xc}$ itself. On the other hand, there also exists a range of physical properties and processes, whose description is substantially influenced by the xc \textit{potential}, $v_\mathrm{xc}\rr := \delta E_\mathrm{xc}[n] / \delta n\rr$. Prominent examples for the latter category are charge-transfer and ionization processes, as well as the description of photoemission spectra. Here, especially the approximate interpretation of Kohn-Sham (KS) eigenvalues as a physically meaningful density of states lies at the focal point of ongoing research (see Ref.~\cite{KronikKuemmel_PES} and references therein).

For applications to real materials, one would wish to have a DFA with a good performance from \emph{both} perspectives. However, this is not the case for many existing DFAs. The development of such a DFA is a long-standing challenge, as discussed below.

One of the exact relations in KS-DFT is the ionization potential (IP) theorem, $-\eps_\mathrm{ho} = I$~\cite{PPLB82,Levy1984,Almbladh1985,Perdew1997a,Harbola99,Dabo14}, which relates the highest occupied (ho) KS eigenvalue, $\eps_\mathrm{ho}$, to the IP, $I$, i.e., the removal energy of one electron from an $N$-electron system.
Consequently, there are two fundamentally different ways to obtain the IP in DFT: by evaluating $-\eps_\mathrm{ho}$ or (with more computational effort) by calculating the energy difference of the ionized and neutral system, with $N_0-1$ and $N_0$ electrons, respectively: $I_{\Delta \textrm{SCF}}=E{(N_0-1)} - E{(N_0)}$. This is usually referred to as the $\Delta$SCF approach. Note that while the latter relies on accurate total \emph{energy} values, the former relies on the \emph{potential} to yield an accurate ho energy level.

The $\Delta$SCF approach, which has been extensively used since the early days of DFT~\cite{TongSham66,MJW,vonBarthHedin73,GunnLund74,JMW75,GunnLund76,GunnJones80,PZ81,Kotochigova97,Kraisler2010,Argaman2013},
usually yields an IP with a satisfactory accuracy of a few percent with respect to experiment, for atoms and small molecules, even with standard (semi-)local DFAs, such as the local spin-density approximation (LSDA)~\cite{Wang1992} or the generalized gradient approximation (GGA)~\cite{Perdew1996}. Comparison of $-\eps_\mathrm{ho}$ to the experimental IP, however, shows poor correspondence for many DFAs. For example, with the aforementioned (semi-)local approximations, one can observe an underestimation of up to 50\%, as manifested in e.g.~\cite{GunnJones80,AllenTozer02,LivshitzBaer07,Korzdorfer2008,SalzBaer09,Klupfel11}. 

This failure has been related to various systematic shortcomings of existing functionals. First, the one-electron self-interaction problem, i.e., the fact that for many DFAs the Hartree energy $E_\mathrm{H}$ is not canceled by $E_\mathrm{xc}$ if evaluated on one-electron ground-state densities~\cite{PZ81}, is well known to have a large impact on the quality of KS eigenvalues~\cite{KronikKuemmel_PES}.
Second, many DFAs show a potential that features an incorrect long-range asymptotic behavior, again with negative effect on the interpretation of KS eigenvalues (see, e.g., \cite{Kronik_JCTC_review12} and references therein). 
Note that while the two issues are related in a physical sense, they are not the same and their connection in the construction of reasonable DFAs is far less obvious~\cite{Schmidt2014a}. Another shortcoming affecting the KS eigenvalues is the deviation of the total energy curve, $E(N)$, as a function of the number of electrons, $N$, from piecewise linearity, for fractional $N$ (see, e.g.,~\cite{Mori-Sanchez2006,Vydrov07,Ruzsinszky2007,Cohen2008,Haunschild2010,Dabo10,Cohen12, SteinKronikBaer_curvatures12,Steinmann13,Borghi14,Dabo14,Mosquera2014a,
Mosquera14a,Nguyen15,Borghi15}). 
In the literature, this phenomenon is sometimes referred to as many-electron self-interaction~\cite{Mori-Sanchez2006,Vydrov07,Ruzsinszky2007,Haunschild2010} or as a (de-)localization error~\cite{Cohen2008,Cohen12,Mosquera14a}. 
For (semi-)local functionals, one obtains a convex energy curve rather than a straight line. Such a deviation reflects negatively also on systems with an integer $N$: it leads to disagreement between the IPs predicted by the $\Delta$SCF method and those predicted via $-\eps_\mathrm{ho}$. This happens because the slope of the energy curve (to the left) equals, according to Janak's theorem~\cite{Janak1978}, the ho KS eigenvalue. Therefore, even when one is interested only in closed systems, with an integer number of electrons, it is important to tackle the problem of lack of piecewise-linearity in order to obtain a physically meaningful value for $\eps_\mathrm{ho}$. 

There exist many approaches to address the aforementioned shortcomings and obtain accurate results for the IP via $\-\eps_\mathrm{ho}$. Self-interaction correction~\cite{PZ81} schemes lead to a significant improvement in the interpretation of KS eigenvalues~\cite{Korzdorfer2009}. Yet, their performance for ground-state energetics is debatable~\cite{Cremer01,Vydrov2004,Vydrov06a,Hofmann12a,Klupfel12}. Approaches that approximate directly the xc potential~\cite{VanLeeuwen1994,Tozer1998,Becke2006,Cencek2013} yield eigenvalues that satisfactorily reproduce the experimental IP, due to modified long-range properties of the potential. However, for these functionals total-energy related quantities are not accessible~\cite{Karolewski2009,GaidukChulkovStaroverov12,Karolewski2013a}.
New types of GGAs can yield significantly improved potential properties~\cite{Armiento2013,Vlcek15,Tran15}, but at the cost of being less accurate for total energies. 
Global hybrid functionals~\cite{Becke1993,Becke1993a,PerErnBurke96,B3LYP_2,AdamoBarone99,ErnzerhofScuseria99}, which linearly combine (semi-)local xc energy components, with a weight $(1-a)$, and exact exchange (EXX, i.e., the Fock-integral evaluated with KS orbitals), with a weight $a$, mitigate the one-electron self-interaction error and often yield an excellent description of properties related to the total energy, for $a \approx 0.25$. However, since the self-interaction is only partly canceled, the KS potential falls off too quickly in the asymptotic limit, and $-\eps_\mathrm{ho}$ is typically far from describing experimental IPs.

Nevertheless, with global hybrids it is possible to find a value of $a$ such that the global hybrid will produce a piecewise linear energy curve, and therefore an improved value for $- \eps_\mathrm{ho}$. This happens because for fractional $N$ the non-local EXX component of the global hybrid produces a concave energy curve (see, e.g.~\cite{Ruzsinszky2007}), while the (semi-)local components usually cause a convex energy curve, which therefore cancel each other. However, this cancellation is achieved with values of $a \approx 0.75$~\cite{Sai11,Imamura11,Atalla13}, which in most cases significantly compromises the performance of the functional for other quantities~\cite{AdamoBarone99,ErnzerhofScuseria99,Zhao08,Marom10,Korzdorfer10,Kronik_JCTC_review12}. 

This creates what we call the `parameter dilemma': while an accurate description of \emph{energy}-related quantities requires a certain value for the functional's parameter, the accurate description of \emph{potential}-related quantities requires a different value, and there is no value that provides a satisfactory description of both~\cite{VermaBartlett12_II,VermaBartlett12_III,VermaBartlett14_IV}. 

Local hybrid functionals~\cite{Cruz1998,Jaramillo2003,Kaupp2007} aim at preserving the good energetics of global hybrid functionals while reducing the self-interaction error by introduction of a more flexible, space-dependent mixing of (semi-)local and non-local components (see~\cite{DeSilva2015} for an overview and discussion). 
However, we recently illustrated using a specially constructed local hybrid functional, termed ISOcc in the following, that the aforementioned `parameter dilemma' persists also there~\cite{Schmidt2014}. Similarly, in range-separated hybrids (RSHs) the values of the range-separation parameter have to be different to accurately reproduce e.g.\ atomization energies and ionization potentials~\cite{LivshitzBaer07}. 

It has been recently shown~\cite{Kraisler2014,Kraisler2013} that an alternative way to improve the prediction of the IP via $-\eps_\mathrm{ho}$ is given by employment of the ensemble approach\red{~\cite{Lieb,PPLB82,LevyPerdew_in_NATO,RvL_adv}} in KS-DFT.
This approach allows for the generalization of the Hartree and xc functionals for fractional $N$ such that the piecewise linearity behavior of the total energy is restored, to a large extent. As a result, better correspondence of $\eps_\mathrm{ho}$ to the experimental IP and to the $\Delta$SCF value is achieved, as demonstrated for the H$_2$ molecule and the C atom with the LSDA.

Here we employ the ensemble generalization procedure proposed in Ref.~\cite{Kraisler2013} to the Hartree and common approximate xc functionals, aiming to address the aforementioned challenge of simultaneous prediction of energy-related and potential-related properties with one DFA. Focusing on the prediction of the IP via the ho KS eigenvalue, we examine a variety of xc approximations: the local spin-density approximation, semi-local generalized gradient approximations as well as global and local hybrids. Results for a representative test set of 26 light diatomic molecules and single atoms are presented. 
We find that the ensemble generalization systematically improves the prediction of the IP, for a wide variety of systems and xc functionals, changing the general tendency from under- to a small overestimation, compared to experiment. This improvement is achieved without any change in total energy-related properties. For hybrids that include a parameter, the IP obtained with ensemble-generalized functionals is found to be only weakly dependent on the parameter value, contrary to common experience with non-generalized hybrids. 
Thus, the ensemble approach eliminates one aspect of the `parameter dilemma'. 

\section{Theoretical background}\label{sec.methodology} %Writer: Eli

For completeness, we briefly present the ensemble generalization to the approximate Hartree-exchange-correlation (Hxc) density functional, focusing on its influence on the highest occupied KS energy level, $\eps_\mathrm{ho}$. A complete derivation can be found in Refs.~\cite{Kraisler2014,Kraisler2013}.

First, we formally consider a system with a fractional number of electrons, $N = N_0 - 1 + \alpha$, where $N_0 \in \mathbb{N}$ and $\alpha \in [0,1]$, so that $\alpha=1$ corresponds to a neutral and $\alpha=0$ to a singly-ionized system. Subsequently, we take the limit $\alpha \rarr 1^-$, focusing on neutral systems with an integer number of electrons. 

\redd{In a landmark article, Perdew {\it et al.} have shown that the zero-temperature ground state of an interacting many-electron %Coulomb 
system possessing a fractional $N$ should be described by an ensemble state~\cite{PPLB82}. This state is a linear combination of the pure ground states for $N_0-1$ and $N_0$ electrons, with the classical statistical weights of $(1-\alpha)$ and $\alpha$, respectively}\footnote{Here and below it is assumed that the ground states of the system of interest and of its ion are not degenerate, or that the degeneracy can be lifted by applying an infinitesimal external field},\footnote{The fact that only contributions from the $N_0-1$- and the $N_0$-states are included relies on the conjecture that the series $E(N_0)$ for $N_0 \in \mathbb{N}$ is a convex, monotonously decreasing series. In other words, all ionization energies $I(N_0):=E(N_0-1)-E(N_0)$ are positive, and higher ionizations are always larger than the lower ones: $I(N_0-1) > I(N_0)$. This conjecture, although strongly supported by experimental data, remains without proof, to the best of our knowledge~\cite{Dreizler1995,Lieb,Cohen12}.}. %In Cohen12, it is in Sec. 4.1.3. 
\redd{The ground-state energy of this ensemble state has then been shown to be equal to $E(N) = (1-\alpha)E(N_0-1) + \alpha E(N_0)$, i.e., it is a piecewise-linear function of $N$~\cite{PPLB82}. This result is a general one, applying to any many-electron system. %ek: added many-
Therefore, in principle it trivially carries over to DFT, because if the {\it exact} exchange-correlation functional is used, DFT-based energies must reproduce the all-electron ones.}

\redd{As mentioned in the introduction, in practice {\it approximate} density functionals often exhibit significant deviations when trying to describe a quantum system with fractional $N$ in KS-DFT, and it has been traditionally assumed that this is just another manifestation of the approximate nature of the functional used. However, in Ref.~\cite{Kraisler2013} it was pointed out that %ek: added these 2 words
much of this deviation is due to the fact that the pure-state exchange-correlation expression is used for both integer and fractional densities, whereas the fractional KS system must itself be in an ensemble state. This happens because %. recall that 
the number of particles in the KS system equals the number of electrons in the real, interacting system, and is also fractional. Therefore, the ground state of the KS system must also be expressed as 
an ensemble of the $(N_0-1)$- and $N_0$- KS states, obtained from the same KS potential, %ek: removed: and this must hold 
even if one uses an approximate functional. Ref.~\cite{Kraisler2013} therefore suggested that any approximate Hartree-exchange-correlation functional can be generalized for an ensemble ground state using ensemble state theory~\cite{Lieb,PPLB82,LevyPerdew_in_NATO,RvL_adv} (for other recent uses of the ensemble approach see Refs.~\cite{GouldDobson2013,Mosquera14a, Goerling15}).} Performing an ensemble average of the many-electron Coulomb operator $\hat W = \half \sum_i \sum_{j \neq i} | \rb_i - \rb_j|^{-1}$ in the KS system, \redd{it has been found that} the pure-state Hxc energy functional can be generalized to ensemble states in the following form:
\begin{equation}\label{eq.ET.gen}
    E_\textrm{e-Hxc}[n^{(\alpha)}] = (1-\alpha) E_\textrm{Hxc}[\rho_{-1}^{(\alpha)}] + \alpha E_\textrm{Hxc}[\rho_0^{(\alpha)}],
\end{equation}
which is exact for the Hartree and exchange components and approximate for the correlation. %ek3: added this line
Here, the index e- indicates that the functional is ensemble-generalized, $E_\textrm{Hxc}$ is the pure-state Hxc functional, $\rho_p^{(\alpha)} \rr$ is defined as the sum of the first $N_0+p$ KS orbitals squared: $\rho_p^{(\alpha)}\rr = \sum_{i=1}^{N_0+p} |\pphi_i^{(\alpha)}\rr|^2$, where $p=-1$ or 0, and $n^{(\alpha)} \rr =(1-\alpha) \rho_{-1}^{(\alpha)} \rr+ \alpha \rho_0^{(\alpha)} \rr$ is the ensemble-state electron density.
\red{When $N$ is an integer, i.e., $\alpha$ assumes the value of 0 or 1, the Hxc energy reduces to that obtained from the underlying pure-state Hxc functional. Therefore, ensemble-generalization does not affect the total energy at integer $N$.}
The generalization in Eq.~(\ref{eq.ET.gen}) is applicable to \emph{any} xc functional and makes the Hartree and the xc energy components \emph{explicitly} linear in $\alpha$. However, there may still remain an \emph{implicit} non-linear dependence of $E_\textrm{e-Hxc}[n^{(\alpha)}]$ on $\alpha$, \red{because the KS orbitals themselves, $\pphi_i^{(\alpha)}\rr$, and consequently $\rho_p^{(\alpha)} \rr$ and $E_\textrm{Hxc}[\rho_p^{(\alpha)}]$, may depend on $\alpha$. The dependence of $\pphi_i^{(\alpha)}\rr$ on $\alpha$ arises from the fact that the KS orbitals are expected to relax as one varies $\alpha$ from 0 (positive ion) to 1 (neutral system) \cite{Kraisler2013,Borghi14, Goerling15}.}

\red{Importantly, Eq.~(\ref{eq.ET.gen}) is derived by considering the generalization of pure-state functionals to ensemble states, without assuming anything {\it a priori} about piecewise-linearity, \redd{because it also applies to approximate exchange-correlation functionals}. Nevertheless, in Refs.~\cite{Kraisler2013,Kraisler2015} it has been shown that by employing Eq.~(\ref{eq.ET.gen}) the energy curve $E(N)$ satisfies the piecewise-linearity criterion much more closely, being slightly concave. The concavity is related to the above mentioned implicit non-linear dependence of the energy on $\alpha$. Another perspective on this approximate piecewise-linearity can be obtained from the fact that Eq.~(\ref{eq.ET.gen}) can be derived, with some further approximations, from different schemes that attempt to enforce piecewise-linearity explicitly~\cite{Borghi14,Dabo14,Borghi15,Nguyen15,Goerling15,Zheng11}.}

Due to the fact that the slope of $E(N)$ changes for all $\alpha$, including $\alpha \rarr 1^-$, it follows from Janak's theorem~\cite{Janak1978}
\footnote{Janak's theorem~\cite{Janak1978} states that the $i$-th KS eigenenergy, $\eps_i$, equals $\de E / \de f_i$ -- the derivative of the total energy of the interacting system, $E$, with respect to the occupation of the $i$-th level, $f_i$. It can be shown that with the exact xc functional the ho eigenenergy, $\eps_\mathrm{ho}$, has to equal $\eps_\mathrm{ho} = \de E / \de \alpha = \de E / \de N = E(N_0) - E(N_0-1) = - I$, i.e., it equals the negative of the IP.}, 
which identifies $\de E / \de N$ with $\eps_\mathrm{ho}$, that the ho energy level has to change, too, even for a system with an integer $N$. This change is obtained in practice from an ensemble generalization of the KS potential, as explained below.

The KS potential is expressed as $v_\textrm{e-\KS} \rr = v_\textrm{ext}\rr + v_\textrm{e-Hxc}[n]\rr$, where $v_\textrm{ext}\rr$ is the external potential and $v_\textrm{e-Hxc}[n]\rr:=\delta E_\textrm{e-Hxc} / \delta n \rr$ is the ensemble-generalized Hxc potential. 
For the limit $\alpha \rarr 1^-$, this potential reduces to a sum of two terms: $v_\textrm{e-Hxc}[n]\rr = v_\textrm{Hxc}[n]\rr + v_0[n]$ -- the usual pure-state Hxc potential,  $v_\textrm{Hxc}[n]\rr$, and a spatially uniform term, $v_0[n]$, which can be written as~\cite{Kraisler2013}
\begin{align}\label{eq.v0}
\nonumber    v_0[n] = E_\textrm{Hxc}[n] &- E_\textrm{Hxc}[n - |\pphi_\mathrm{ho}|^2] \\
&- \int  |\pphi_\mathrm{ho} \rr|^2 v_\textrm{Hxc}[n] \rr d^3r.
\end{align}
Here and below the superscript $(\alpha)$ is dropped at the limit $\alpha \rarr 1^-$ for brevity. \red{Note that the ensemble-generalized KS potential does not vanish at $r \rarr \infty$, but asymptotically approaches $v_0[n]$.}
We stress that $v_0[n]$ is a well-defined, rather than arbitrary, potential shift. It \emph{must} be taken into account for the ensemble-generalized functional in order for the ho KS eigenvalue to equal $\de E/ \de N$, i.e., to obey Janak's theorem. \red{Note that the shift discussed here is different from the one recently proposed by Zahariev and Levy~\cite{ZaharievLevy14}. As clarified in Ref.~\cite{Goerling15}, in Ref.~\cite{ZaharievLevy14} the potential shift makes the energy of the KS system equal the energy of the interacting system. Here, however, the shift emerges naturally from the ensemble treatment and is essential to obtaining results that are consistent with Janak's theorem.}
Also note that while the result above has been presented in a spin-independent form for simplicity, in practice, in spin-dependent calculations, there exist potential shifts $v_0^\s$ to both spin channels $\s = \up, \dn$. Calculating $v_0^\s$ with Eq.~(\ref{eq.v0}), we take the ho level to be the highest occupied level in the $\s$-channel considered (noted as $\s-$ho). In the following, however, if not stated explicitly otherwise, when mentioning the ho level we refer to the global ho: $\eps_\mathrm{ho}=\max_\s \eps_\mathrm{ho}^\s$, i.e.\ the one of the two $\s$-ho levels which is higher in energy; the same applies for the ensemble-generalized ho level, $\eps_\textrm{e-ho}$.

To summarize, as a result of the approximate ensemble generalization of the Hxc functional (Eq.~(\ref{eq.ET.gen}))~\cite{Kraisler2013, Kraisler2014}, in the limit of integer $N$ the KS potentials exhibit spatially uniform shifts $v_0^\s$, such that all KS eigenvalues of the same spin channel are shifted by the same value \red{(see, e.g., Fig.\ 4 in Ref.~\cite{Kraisler2015}. The KS orbitals, and as a result the density and the total energy, are not changed and remain the same as those obtained with the underlying Hxc functional. Furthermore, because all eigenvalues are shifted by the same amount, eigenvalue differences (as well as quantities based on them, e.g., in linear response time-dependent DFT \cite{Casida95}) are not affected either.} Therefore, the $\s-$ho energy levels of the ensemble-generalized functional can be expressed as $\eps_\textrm{e-ho}^\s = \eps_\mathrm{ho}^\s + v_0^\s$,  being a sum of the $\s-$ho level that emerges from a standard KS-DFT calculation prior to the ensemble generalization and the potential shift of the relevant spin channel, calculated according to Eq.~(\ref{eq.v0}). 
Comparing both $\eps_\textrm{e-ho}$ and $\eps_\mathrm{ho}$ to experimental IPs and $-I_{\Delta \textrm{SCF}}$ is the main subject of Sec.~\ref{sec.results}.

\section{Computational details}\label{sec.numdet}
We concentrate on a relatively elementary, yet chemically representative, set of systems, consisting of 18 light diatomic molecules: H$_2$, LiH, Li$_2$, LiF, BeH, BH, BO, BF, CH, CN, CO, NH, N$_2$, NO, OH, O$_2$, FH, F$_2$, and their 8 constituent atoms. The simplicity of the systems allows us to keep computational costs low and to refrain from introducing additional sources of error, e.g., searching for an optimal geometry in systems with many degrees of freedom. At the same time, systems of single-, double-, and triple-bond molecules as well as atoms (no bonding) are included in the test set, which makes the set representative of more complicated systems, as shown in previous work (see, e.g.,~\cite{Perdew1996,Schmidt2014}).

All calculations were performed using the program package \darsec~\cite{Makmal2009,Makmala}, an all-electron code that allows for electronic structure calculations of single atoms or diatomic molecules on a real-space grid represented by prolate-spheroidal coordinates. \darsec \, allows one to solve the KS equations self-consistently for density- as well as orbital-dependent functionals. For the latter, a local, multiplicative xc potential is obtained by employing the KLI~\cite{Li1992a} approximation to the optimized effective potential (OEP)~\cite{Grabo1997,Engel2011,Kummel2008} formalism.  
Use of this approximation has been justified in Ref.~\cite{Li1993} for the EXX functional and in Ref.~\cite{Schmidt2014} for the ISOcc local hybrid functional.

For all systems, an accuracy of 0.0005 \hartree \, in the total energy and in the ho KS eigenvalue has been achieved by appropriately choosing the parameters of the real-space grid and by iterating the self-consistent DFT cycle. For molecules the bond length was taken from experiment~\cite{HandChemPhys92,webbook}. 
Differences due to atomic relaxation were found to be insignificant\footnote{For the LSDA, relaxation runs have been performed for all molecules. It was found that the experimental bond length lies within the numerical error range of the relaxed bond length in all cases. 
We checked that the ho and $e-$ho energy values for the relaxed geometries agree with the ones at experimental geometries within 0.002 \hartree, except for H$_2$, NH and F$_2$, where the difference reaches 0.005 \hartree. For the ISOcc functional similar relaxation checks were performed, as described in Ref.~\cite{Schmidt2014}},\footnote{In this context, we note that the values reported by some of us in Ref.~\cite{Kraisler2013} for the relaxed H$_2$ molecule, namely the \emph{e}-ho energy, $\eps_\textrm{e-ho}$, and as a result -- the fundamental gap of the ion, $E_\mathrm{g}$, are slightly different upon closer observation. In fact, at the relaxed bond length of $L=1.45$ Bohr, these values are $\eps_\textrm{e-ho} = 0.618$ \hartree $ = 1.236$ Ry and $E_\mathrm{g} = 0.671$ \hartree $ = 1.341$ Ry and not 1.223 Ry and 1.320 Ry, respectively. The difference originates from retrieving the value for $\eps_\textrm{e-ho}$ directly, not relying on the chemical potential $\mu$ calculated in the \darsec \: program, with a temperature of 1K.}.
\redd{The net spin of the neutral systems was also taken to be as in experiment. The spin configurations of cations (used below for calculating ionization potentials from total energy differences) was obtained by removing an electron from the highest occupied orbital of the neutral.}

\section{Results}\label{sec.results}

\subsection{Effect of the ensemble correction - O$_2$ as a prototypical case}\label{subsection.O2}
%1) Writer: Tobi. Diagram: Eli.

\red{Previous work~\cite{Kraisler2013,Kraisler2015} has already demonstrated that the ensemble generalization of Eq.\ (\ref{eq.ET.gen}) significantly reduces the deviation from the piecewise-linearity condition for the total energy, i.e., greatly diminishes the delocalization error, and as a consequence eliminates the fractional dissociation error in diatomic molecules. Here we focus on the potential shifts (Eq.\ (\ref{eq.v0})) that emerge from the ensemble-generalization and their effect on the Kohn-Sham energy levels. In particular, we consider the prediction of the IP via $\eps_\textrm{e-ho}$.} 

For a clear understanding of the results presented in this paper, it is of advantage to first illustrate the effect of the potential shift mechanism, given by Eq.~(\ref{eq.v0}), on the eigenvalue structure of a particular system with an integer number of electrons. Here we provide a detailed presentation of a selected system - the O$_2$ molecule, computed with the Perdew-Burke-Ernzerhof (PBE) GGA~\cite{Perdew1996} at its experimental bond length of $2.2819 \ \mathrm{bohr}$.

Due to its electronic ground-state configuration, $ ^3 \Sigma^-_g$, this system must 
be treated in a spin-polarized formalism. Consequently, it provides an interesting example %ek: old: to study 
for how eigenvalues belonging to different spin channels are shifted when the corresponding ensemble potential shift, $v^{\s}_0$, is applied. 
 
\begin{figure}[h]
  \includegraphics[width=.45\textwidth]{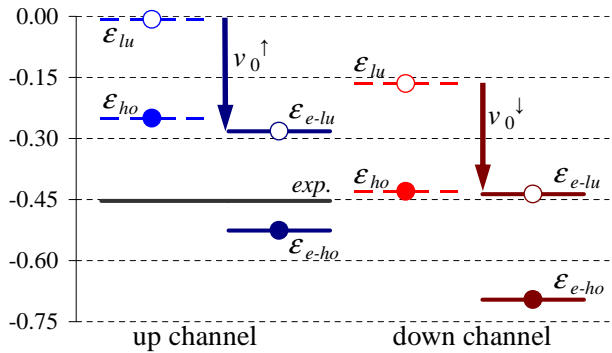}\\
  \caption{Diagram of the highest occupied and lowest unoccupied KS-PBE eigenvalues of the O$_2$ molecule, for both spin channels, before and after applying the potential shifts of Eq.~(\ref{eq.v0}), along with the negative of the experimental IP. All values are in \hartree.}
\label{fig.O2}
\end{figure}

For this purpose, the positions of the highest occupied ($\eps_\mathrm{ho}^\s$) and 
lowest unoccupied ($\eps_\mathrm{lu}^\s$) KS eigenvalues for both spin channels are depicted in Fig.~\ref{fig.O2}. The eigenvalues changed by the respective potential shift, i.e., $\eps_\textrm{e-ho}^\s = \eps_\mathrm{ho}^\s + v_0^\s$ and $\eps_\textrm{e-lu}^\s = \eps_\mathrm{lu}^\s + v_0^\s$, as well as the negative of the experimental IP, $-I_\textrm{exp}$, are also included in the figure.

It can be readily observed that the unshifted highest occupied eigenvalue of the up channel, $\eps^\up_\mathrm{ho} = -0.251$ \hartree, which lies higher than its spin down counterpart, poorly reproduces the negative of the experimental IP of the O$_2$ molecule. In fact, with PBE it underestimates the experimental IP of $I_\textrm{exp} = 0.453$ \hartree~\cite{webbook} by $45\ \%$, a value that is quite typical for other systems as well. However, after application of the potential shift, the highest occupied eigenvalue is $\eps^\up_\textrm{e-ho} = -0.526$ \hartree, i.e., the experimental IP is now \textit{over}estimated by $16\ \%$. As shown below, this is a typical result also for other systems \textit{and} other functionals. 

From the results presented for the ensemble generalized PBE functional, the question of how other DFAs perform for the same system naturally arises. In particular, the change of the eigenvalues obtained with functionals containing a varying amount of non-local EXX is of great interest, as we know that a greater percentage of EXX already leads to a more accurate description of IPs via the highest occupied eigenvalue.

\begin{figure}[h]
  \includegraphics[scale=0.45,angle=0]{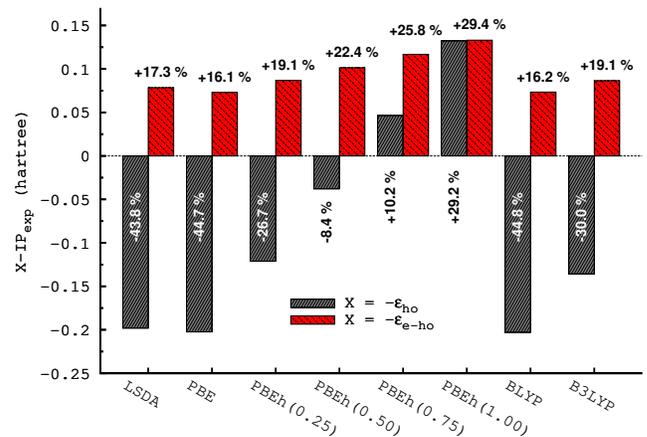}\\
  \caption{Comparison of $-\eps_\mathrm{ho}$ and $-\eps_\textrm{e-ho}$ to the experimental IP of the O$_2$ molecule, calculated with different DFAs. The corresponding labels provide the relative deviation in percent.}
\label{fig.O2_funcs}
\end{figure}

Fig.~\ref{fig.O2_funcs} provides a comparison of the unshifted and shifted ho eigenvalues to the experimental IP for O$_2$.  Besides (semi-)local functionals such as the LSDA, PBE and BLYP~\cite{Becke1988,Lee1988,Miehlich1989}, we also ensemble-generalized the global hybrid functionals B3LYP~\cite{B3LYP_2} and PBEh($a$)~\cite{AdamoBarone99} (employed within the KS scheme using the KLI approximation), with $a$ denoting the fixed amount of EXX combined with $(1-a)$ of PBE exchange and with full PBE correlation.

The unshifted eigenvalues for the three purely (semi-)local functionals (LSDA, PBE, BLYP) underestimate the IP by $\sim 45\ \%$. After ensemble generalization, we observe an overestimation by $\sim 16\ \%$. It is instructive to check to which extent this overestimation comes from errors in calculating total energy differences that are inherent to the underlying functional, and to which extent they arise from the ensemble generalization process~\footnote{We recall that the latter does not produce a strictly piecewise linear energy curve $E(N)$, but there typically remains some concavity, which is attributed to the implicit dependence of $E(N)$ on $\alpha$ via the KS orbitals. This concavity affects the value of $\eps_\textrm{e-ho}$. However, even in case $E(N)$ would be exactly piecewise linear, $\eps_\textrm{e-ho}$ would reproduce $- I_{\Delta \textrm{SCF}}$, rather that the experimental IP.}. Therefore, we compare the shifted and unshifted eigenvalues to $- I_{\Delta \textrm{SCF}}$. For O$_2$ computed with PBE, one obtains $I_{\Delta \textrm{SCF}} = 0.464$ \hartree, which deviates from experiment by only 2.3 \%. With respect to this quantity, the unshifted eigenvalue yields an underestimate of $46\ \%$, while the shifted value overestimates it by $14\ \%$. We therefore realize that most of the discrepancy comes from the concavity that remains in the $E(N)$ curve even after ensemble generalization.

In the global hybrid, PBEh$(a)$, increasing the intrinsic amount of EXX significantly improves the correspondence of the unshifted eigenvalue to experiment. Due to the fact that the KS potential decays asymptotically more slowly with a growing value of $a$, the IP via $-\eps_\mathrm{ho}$ is very sensitive to non-local functional components included. Changing from under- to overestimation, an optimal description of $I_\textrm{exp}$ is reached for this system with $a\approx 0.6$. However, for ensemble generalized DFAs the value of $-\eps_\textrm{e-ho}$ systematically overestimates the IP with respect to experiment for the O$_2$ molecule, regardless of the value of $a$, while at the same time being far less sensitive to the amount of non-locality in the functional expression. While for "plain" PBE the relative error now reads $\sim +16\ \%$, it increases to $+29\ \%$ when full non-local exchange combined with PBE correlation is used. The reason for this reduced sensitivity lies in the following mechanism: while the absolute value of $\eps_\mathrm{ho}$ grows with increasing $a$, the potential shift $v_0$ is reduced, roughly commensurately, because the Hartree+EXX functional has zero potential shift~\cite{Kraisler2013}. 

%2) Writer: Tobi.
\subsection{Evaluating the test set - a systematic study} 

Following the illustration of the mechanism of the potential shift for a single system, we now focus on the mean discrepancy in the evaluation of the experimental IP via shifted and unshifted KS eigenvalues, for a variety of functionals (\red{see supplemental material at [URL will be inserted by AIP] for detailed numerical data in tabular form}). We use the test set of systems introduced in Sec.~\ref{sec.numdet} as a basis for averaging.

We emphasize that the eigenvalues shift is expected to improve the correspondence between the negative of the ho eigenvalue and the ionization energy obtained via the $\Delta$SCF method, for a given DFA. We compare the shifted and unshifted eigenvalues to experiment, and not to $\Delta$SCF values, relying on the aforementioned fact that the $\Delta$SCF reliably describes systems of our test set, with small average relative errors: 3.4 \% for PBE and 4.2 \% for both the LSDA and ISOcc$(c=0.5)$. 

We define the averaged relative error in the ionization potential    
\begin{equation}\label{eq.delta}
    \delta_\textrm{IP} = \sqrt{ \frac{1}{M} \sum_{j=1}^M \lp( \frac{-\eps^{(j)} - I^{(j)}_\textrm{exp}}{I^{(j)}_\textrm{exp}} \rp)^2}.
\end{equation}
Here, the index $j$ runs over all systems in the test set up to the total number $M = 26$, and $\eps$ stands either for the shifted ($\eps_\textrm{e-ho}$) or unshifted ($\eps_\mathrm{ho}$) highest occupied KS eigenvalue.

Note that in Eq.~(\ref{eq.delta}) the unsigned deviation from experimental IPs is employed
to avoid a misleading result of zero average relative error, which emerges when there occurs an overestimation for some systems and an underestimation for others. 
However, in order to be able to distinguish between systematic over- or underestimation, an additional measure is defined accordingly:  
\begin{equation}\label{eq.sigma}
    S = \frac{1}{M} \sum_{j=1}^M \sgn \lp( -\eps^{(j)} - I^{(j)}_\textrm{exp} \rp).
\end{equation}
While $\delta_\textrm{IP}$ provides the mean deviation from experimental values in \%, the quantity $S$ indicates the average trend of the prediction, being naturally confined to the interval $[-1,1]$. 
Namely, for a systematic overestimation we obtain $S = 1$, and for a systematic underestimation $S = -1$. Both quantities, $\delta_\textrm{IP}$ and $S$, were obtained for various DFAs and their ensemble-generalized counterparts. 

\begin{figure}[h]
  \includegraphics[scale=0.45,angle=0]{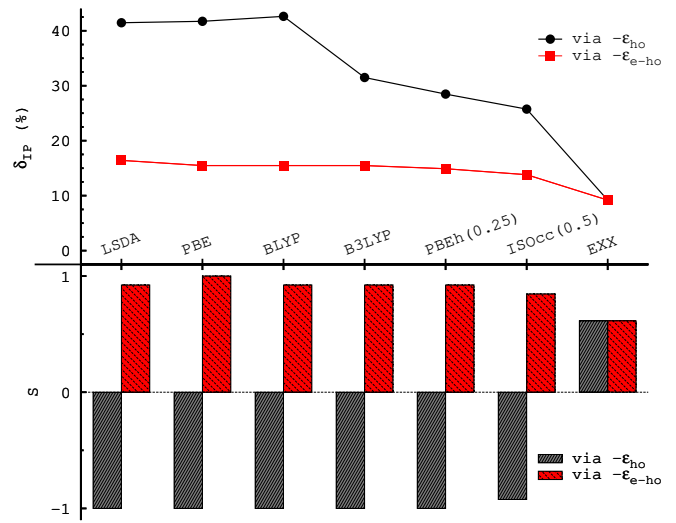}\\
  \caption{Average relative error $\delta_\textrm{IP}$ in $\%$ (upper axis) 
  and signum function $S$ (lower axis) for the LSDA, PBE, BLYP, B3LYP, PBE$(a=0.25)$, ISOcc$(c=0.5)$ and pure EXX (black) as well as their ensemble-generalized versions (red).}  
\label{fig.varfuncs}
\end{figure}

\begin{figure}[h]
  \includegraphics[scale=0.45]{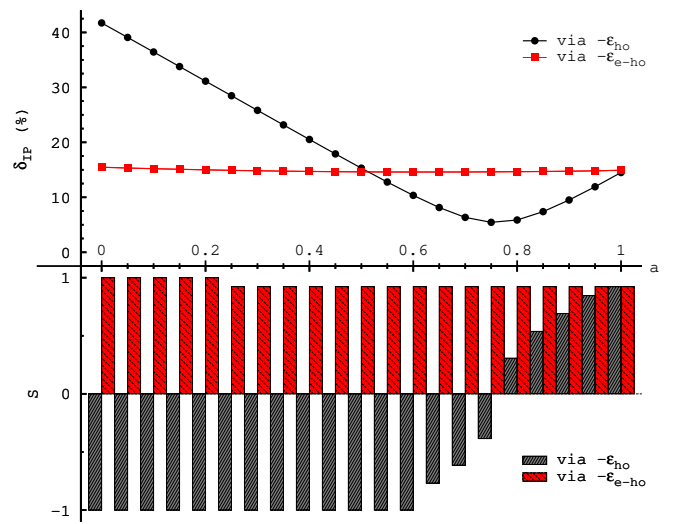}\\
\caption{Average relative error $\delta_\textrm{IP}(a)$ in $\%$ (upper axis) 
and signum function $S(a)$ (lower axis) 
for PBEh$(a)$ (black) and e-PBEh$(a)$ (red) in dependency on the parameter $a$.}
\label{fig.pbeh}
\end{figure}

\begin{figure}[h]
  \includegraphics[scale=0.45]{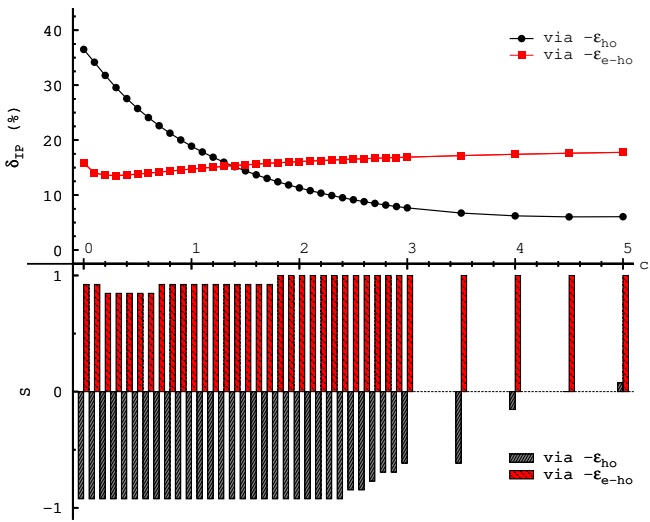}\\
  \caption{Average relative error $\delta_\textrm{IP}(c)$ in $\%$ (upper axis) 
  and signum function $S(c)$ (lower axis) 
  for ISOcc$(c)$ (black) and e-ISOcc$(c)$ (red) in dependency on the parameter $c$.}
\label{fig.isocc}
\end{figure}

Fig.~\ref{fig.varfuncs} shows the corresponding results for the LSDA, the semi-local PBE and BLYP, the global hybrid functionals B3LYP and PBEh(0.25), the EXX, and the ISOcc(0.5) local hybrid functional. 
\red{Note that for EXX the results for the regular and ensemble-generalized functional coincide, because the Hartree+EXX functional exhibits a zero potential shift~\cite{Kraisler2013}}.
\footnote{Note that \red{the combination of the} EXX \red{functional with the standard form for the Hartree functional results in an} intrinsically ensemble-generalized \red{functional if} the ground state is described by an ensemble comprised of \emph{two} pure \red{many-electron states. This is the case throughout this work as we describe the ionization process by extracting an electron from a specific spin-channel.} If the number of \red{many-electron} states is larger than two \red{(as is the case, e.g., if both spin channels are fractionally occupied), then the EXX is \emph{not} intrinsically ensemble-generalized, but an appropriate ensemble generalization, proposed in Ref.~\cite{GouldDobson2013}, is available.}}
Fig.~\ref{fig.pbeh} provides the corresponding results for the PBEh$(a)$ global hybrid functional as a function of the parameter $a$, i.e., on various amounts of non-local EXX
~\footnote{When calculating the NH molecule with the LSDA or PBEh$(a)$ using values of $0\le a \lesssim 0.55$, the global $\eps_\mathrm{ho}$ and $\eps_\textrm{e-ho}$ do not belong to the same spin channel, a behavior that has not been observed in any other system in our test set.}.
The value $a=0$ in this figure reproduces the PBE result.
Fig.~\ref{fig.isocc} depicts $\delta_\textrm{IP}$ and $S$ obtained with the local hybrid ISOcc. The latter functional was developed using the so-called local mixing function, rather than a fixed mixing ratio of non-local and semi-local components. It contains a free parameter $c$, which implicitly determines the intrinsic amount of EXX included in the local hybrid. Higher $c$ values correspond to a higher fraction of EXX being included. Consequently, for ISOcc both the quantities given by Eqs.~(\ref{eq.delta}) and (\ref{eq.sigma}) are functions of this parameter $c$, in analogy to the global hybrid PBEh$(a)$. 

In principle, both parameters $a$ and $c$, in PBEh$(a)$ and ISOcc$(c)$, respectively, are free. However, as mentioned earlier, it is known that in terms of total energy related quantities PBEh$(a)$ performs best for $a=0.25$, while we recently showed that for the ISOcc$(c)$ functional the optimal parameter value is $c=0.5$. Therefore, both functionals, using their optimal respective parameters play a special role in the following discussion, and Fig.~\ref{fig.varfuncs} shows their performance in comparison to the other DFAs.  

Fig.~\ref{fig.varfuncs} clearly indicates that using the unshifted eigenvalues $\eps_\mathrm{ho}$, the three (semi-)local functionals LSDA, PBE, and BLYP strongly and systematically underestimate $\delta_\textrm{IP}$ by $\approx 41-43 \%$. Regarding hybrids, for the global hybrid B3LYP we obtain an underestimation of $31 \%$, for PBEh$(a=0.25)$ $28 \%$ and for the local hybrid ISOcc$(c=0.5)$ $26 \%$.  
The improvement of hybrids over (semi-)local functionals is explained by the fact that the non-local terms in hybrids lead to a partial cancellation of the self-interaction error and an improved behavior of the xc potential in the asymptotic limit.

If the parameters $a$ and $c$ are varied, Figs.~\ref{fig.pbeh} and~\ref{fig.isocc} illustrate that when using $\eps_\mathrm{ho}$ the global hybrid PBEh$(a)$ and the local hybrid ISOcc$(c)$ show a transition in their parameter-dependent $S$-function from negative to positive values. This feature clearly indicates that, for the systems studied here, it is possible to fit the corresponding functional parameter for a given system so that $\eps_\mathrm{ho}$ exactly gives the experimental IP. 
If $a$ and $c$ are optimized to reduce the error $\delta_\textrm{IP}$, we obtain an underestimation of 5 \% for $a=0.75$ in PBEh and of 6 \% for $c=4.5$ in ISOcc. Therefore, by changing the parameters $a$ and $c$, we are able to strongly reduce the average error in the IP of our test set. However, this comes at a price in total energy-related quantities, as has been shown in Refs.~\cite{Sai11,Schmidt2014}, and is the subject of the so-called `parameter dilemma' presented in Sec.~\ref{sec.intro}.

When using the ensemble-corrected highest occupied eigenvalues $\eps_\textrm{e-ho}$, we obtain a completely different picture. First, the systematic underestimation now changes to an overestimation. All of the aforementioned functionals now show a very similar average error of $\delta_\textrm{IP} \approx 14-17 \%$ , which is significantly smaller than the results from non-generalized functionals. Second, for ensemble-generalized hybrid functionals e-PBEh$(a)$ and e-ISOcc$(c)$, there is no transition from an underestimation to overestimation regime, but rather a systematic overestimation of the IP, independent of the parameter value. In other words, the amount of non-locality included in the hybrid functional plays a minor role in the description of IPs via shifted KS eigenvalues, in contrast to their unshifted counterparts. This confirms that the mechanism of cancellation between the change in the potential shift of Eq.~(\ref{eq.v0}) and the highest occupied eigenvalue with a varying amount of non-locality is not particular to the O$_2$ molecule, but rather a systematic feature of ensemble-generalized functionals. Furthermore, in the ensemble-generalized version of DFAs the `parameter dilemma' does not emerge: since the ensemble-generalized eigenvalues describe IPs with an accuracy almost independent of the amount of EXX included, one cannot deduce a preferred value of the parameter by minimizing $\delta_\textrm{IP}$. Therefore, in principle one could use the functional with the parameter optimized to describe binding processes and structural quantities, and rely on the description of IPs via the shifted eigenvalues $\eps_\textrm{e-ho}$. In this case, our results for $\delta_\textrm{IP}$ using thermochemically optimized functionals with ensemble-generalization (such as e-B3LYP, e-PBEh$(a=0.25)$ and e-ISOcc$(c=0.5)$) indicate a clear improvement over their non-generalized counterparts. 

Our results further indicate that even functionals whose xc terms were constructed on different grounds and from different perspectives, such as for example the PBE and BLYP functional, yield similar values of roughly $\delta_\textrm{IP}\approx 15\ \%$ after applying the ensemble generalization. 
As even the inclusion of non-local components does not lead to significant change, one might wonder if this "natural border" of 15\ \% is inherent to the ensemble shift mechanism \textit{regardless} of the specific form of the respective DFA put to task. This question has been checked by varying the parameters $\mu$ and $\kappa$ used in the construction of the PBE exchange functional~\cite{Perdew1996,Pedroza2009}. We find that for different choices of $\mu$ and $\kappa$ one obtains different values for the average relative error $\delta_\textrm{IP}$. 
For instance, using PBE exchange with a value of $\mu=1.0$ together with the original $\kappa = 0.8401$ results in an error of $\delta_\textrm{IP} =20 \%$ when using $\eps_\textrm{e-ho}$, while a combination of the original $\mu=0.21951$ and $\kappa=5.0$ leads to $\delta_\textrm{IP} =8 \%$. From this we conclude that the ensemble-generalization as such does not lead to a fixed systematic error in the description of experimental IPs via KS eigenvalues. However, the results of this subsection suggest that after the ensemble generalization the functionals examined here have a common missing part, which causes the described discrepancy in $\delta_\textrm{IP}$.

\red{
Before concluding this sub-section, we note that while we have focused our work on the IP of neutral atoms and diatomic molecules, IPs of ions may in principle be assessed in the same manner. In particular, the electron affinity (EA) of the neutral can be explored as the IP of the singly charged anion (barring geometrical relaxation). Unfortunately, for the atoms and very small molecules studied here, it is well-known~\cite{Engel_in_Primer,KimSimBurke11,Guliamov07} that with common semi-local approximations negative ions of small systems may erroneously be predicted to be unstable. However, when performing calculations with finite basis sets, as in, 
e.g.,~\cite{GalbraithSchaefer96}, unbound states can be artificially stabilized~\cite{RoeschTrickey97}, because the basis set effectively confines the unbound electron to the vicinity of the neutral system.  Because the ensemble generalization discussed here does not change the total energies of systems with integer electrons (including neutrals, cations, and anions), anions that are not bound with the underlying xc functional will remain unbound even if its ensemble-generalized version is employed. Furthermore, although ensemble-generalization will generally shift the energy levels, including the unoccupied ones, the question of whether the lowest unoccupied KS orbital has a bound or unbound character will not be affected \cite{Kraisler2015}, because orbitals are unchanged by a uniform shift of the potential.}
 
%3) Writer: Eli
\subsection{Ensemble-generalization and the \emph{Aufbau} principle} \label{sec.Aufbau}

In general, at zero temperature the energy levels in the KS system have to be occupied according to the \emph{Aufbau} principle, i.e., the levels are occupied without 'holes', starting with the lowest ones up. In the following, we term such an occupation \emph{proper}. An example for a proper occupation is given in Fig.~\ref{fig.O2} for the O$_2$ molecule.  	All calculations performed for this work, except for a few discussed below, yield proper occupation.

In spin-polarized calculations a special situation can occur, when each of the spin channels is occupied properly itself, while the system as a whole possesses a 'hole' in its occupation. For example, this happens when the lu level of the $\dn$-channel appears lower than the ho level of the $\up$-channel. An occupation of this kind is termed \emph{proper in a broad sense}. It is emphasized here that a broad-sense-proper density obeys all the required restrictions related to a rigorous definition and differentiability of energy functionals~\cite{Kraisler2009,Kraisler_thesis}, therefore it can serve as a legitimate solution of a many-electron system. In the past, broad-sense-proper occupations have been observed in certain transition-metal and lanthanide atoms and ions in LSDA and PBE calculations~\cite{Kraisler2010}, as well as in the Li atom with the EXX~\cite{Makmal2011} and with the exact KS potential, which has been obtained from accurate wave-function-methods based spin densities~\cite{Gritsenko2004}. The latter result strengthens our understanding that a broad-sense-proper occupation is not necessarily an artifact of some DFAs, but rather is an expected result, because it may appear even with the exact functional. 

\begin{figure}[h]
  \includegraphics[width=.45\textwidth]{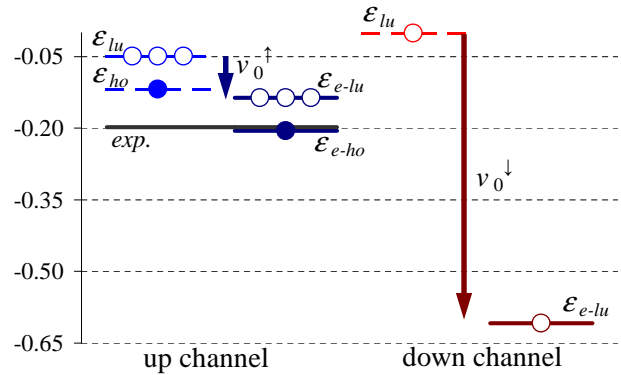}\\
  \caption{Diagram of the highest occupied and lowest unoccupied KS eigenvalues of the Li atom, for both spin channels, before and after application of the potential shifts of Eq.~(\ref{eq.v0}), obtained  within the PBE functional, along with the negative of the experimental IP. All values are in \hartree. The highest occupied eigenvalue at the $\dn$-channel is lower than -0.65 \hartree, and is therefore not shown for clarity.}
\label{fig.Li}
\end{figure}

In the current work we find that the ensemble generalization, by means of the potential shifts $v_0^\s$, yields broad-sense-proper results for systems which appeared strictly proper before. Figure~\ref{fig.Li} illustrates the situation for the Li atom calculated with PBE and e-PBE. Due to the fact that $v_0^\up = -0.087$ \hartree, while $\red{v_0^\dn} = -0.603$ \hartree, the $\dn$-e-lu level appears below the $\up$-e-ho level, causing a broad-sense-proper occupation. The significant differences in the values of the two potential shifts are associated with the different nature of the $\s$-ho orbitals: the $\up$-ho orbital is a relatively delocalized, high-lying 2$s$ orbital, whereas the $\dn$-ho orbital is a localized, low-lying 1$s$ orbital. A similar situation is observed in the Na atom, for which we obtained a broad-sense-proper result, too. To summarize, we view the appearance of broad-sense-proper occupations in the ensemble treatment of  the alkaline atoms Li and Na as another feature of the exact DFT result, which has been recovered by the ensemble generalization.

\subsection{The derivative discontinuity in ensemble-generalized functionals} %Writer: Eli

While in the current work we are concerned primarily with the IP of atoms and molecules, it is worth discussing a related quantity -- the (fundamental) gap, $E_\mathrm{g}$. By definition, $E_\mathrm{g} = I-A$, where $A$ is the electron affinity, i.e., the energy  gained by adding one electron to the system. As opposed to the IP, the gap of the interacting system does not equal the gap of the KS system, $E_\mathrm{g}^\textrm{\KS} = \eps_\mathrm{lu} - \eps_\mathrm{ho}$, even for the exact xc functional. Instead, $E_\mathrm{g} = E_\mathrm{g}^\textrm{\KS} + \Delta$, where $\Delta$ is the derivative discontinuity (DD) -- a ''jump'' experienced by the KS potential when it is varied with respect to $N$, and $N$ crosses an integer value~\cite{PPLB82,Perdew1983,Perdew_in_NATO,Godby87,Godby88,Harbola98,Chan99,AllenTozer02,Teale08}. 

There exist several ways to find the DD. First, for finite systems it can be obtained using total energy differences:\begin{equation}
\Delta_\mathrm{E} = E(N_0+1) - 2E(N_0) + E(N_0-1) - E_\mathrm{g}^\mathrm{\KS}.
\end{equation}
Second, the DD can be obtained as suggested in~\cite{KLI92a,Gorling1995}:
\begin{equation}
\Delta_\textrm{OEP} = \sandw{\pphi_\mathrm{lu}}{u_\textrm{xc,lu}}{\pphi_\mathrm{lu}} - \sandw{\pphi_\mathrm{lu}}{v_\textrm{xc}}{\pphi_\mathrm{lu}}
\end{equation}
where $u_{\textrm{xc},i} \rr := \pphi_i^{-1}\rr \delta E_\textrm{xc} / \delta \pphi_i\rr$, i.e., the orbital-specific xc potential of the $i$-th orbital and $v_\textrm{xc}:=\delta E_\textrm{xc} / \delta n$, i.e., the local xc potential, which in general has to be obtained via the OEP procedure~\cite{Grabo1997,Engel2011,Kummel2008} (hence the index OEP). The derivation of $\Delta_\textrm{OEP}$ assumes the "alignment equality" $\sandw{\pphi_\mathrm{ho}}{u_\textrm{xc,ho}}{\pphi_\mathrm{ho}} = \sandw{\pphi_\mathrm{ho}}{v_\textrm{xc}}{\pphi_\mathrm{ho}}$, which determines the free constant in $v_\textrm{xc} \rr$ as part of the OEP procedure.
Additional approaches to introduce the derivative discontinuity include Refs.~\cite{Kohn86,Andrade11,Gidopoulos12, Chai2013,Baerends2013} (see also~\cite{Cohen12,Dabo13_PsiK,Kraisler2014,Mosquera14a} for an overview). 

Finally, an approximation for the DD can be obtained from an ensemble treatment for a given underlying Hxc functional, as proposed in Ref.~\cite{Kraisler2014}: 
\begin{align}\label{eq.Delta}
\nonumber \Delta_\textrm{ens} &= E_\textrm{Hxc}[n + |\pphi_\mathrm{lu}|^2] - 2E_\textrm{Hxc}[n] + E_\textrm{Hxc}[n - |\pphi_\mathrm{ho}|^2] + \\ 
&+ \int  d^3r \,\, v_\textrm{Hxc}[n]\rr \lp( |\pphi_\mathrm{ho}\rr|^2 - |\pphi_\mathrm{lu}\rr|^2 \rp)
\end{align}

The first way requires three independent self-consistent calculations of the total energy (hence the index E): of the neutral system, the cation and the anion. In contrast, the second and third ways yield the DD from KS quantities of the neutral system only, which is an advantage when considering infinite systems.

Relying on our experience with ensemble-generalized calculations for atoms and small molecules (Ref.~\cite{Kraisler2013} and this work), we expect $\Delta_\textrm{ens}$ obtained with an approximate xc functional to be larger than $\Delta_\mathrm{E}$. As has been shown in Fig.~2 of Ref.~\cite{Kraisler2014} (lower panel), $\eps_\textrm{e-ho}$ is obtained as being somewhat too low immediately to the left of an integer $N$ and somewhat too high immediately to the right of it. As a result, $\Delta_\textrm{ens}$ overestimates the true discontinuity. This overestimate is related to the residual concavity of the $E(N)$ curve after the ensemble generalization. In the current study, we showed that 
the overestimate in $\eps_\textrm{e-ho}$ to the left of the integer point, which corresponds to the negative of the IP, is systematic, i.e., it happens in various systems and with different functionals. Consequently, we expect a systematic overestimate for $\Delta_\textrm{ens}$ and the resulting $E_\mathrm{g}$. 

The discrepancy between $\Delta_\textrm{OEP}$ and $\Delta_\textrm{ens}$ has a different origin. While $\Delta_\textrm{OEP}$ originates because the KS potentials are differently "aligned" (see above) to the left and to the right of an integer point, $\Delta_\textrm{ens}$ comes from two sources (see Ref.~\cite{Kraisler2014} for detailed explanations): 
the first is the same as for $\Delta_\textrm{OEP}$; the second is the fact that the ensemble-generalized KS potential does not approach zero at $r \rarr \infty$, but rather a constant $v_0$ (see Eq.~(\ref{eq.v0})), which is different to the left and to the right of an integer $N$.
$\Delta_\textrm{OEP}$ does not consider the second source described above, assuming (correctly in the context of Refs.~\cite{KLI92a,Gorling1995}) that the potentials asymptotically tend to zero. In fact, it can be analytically shown that $\Delta_\textrm{OEP}$ is an ingredient in $\Delta_\textrm{ens}$, which was denoted by $\Delta_1$ in Ref.~\cite{Kraisler2014}.

\section{Conclusions and Summary}\label{sec.conclusions} %Writer: not assigned yet.

In the current work, we employed the ensemble-generalization procedure~\cite{Kraisler2013} for a test set of 26 diatomic molecules and single atoms, for a variety of xc functionals. These include the local spin-density approximation (LSDA), the semi-local PBE and BLYP, the global hybrids B3LYP and PBEh(a), and the local hybrid ISOcc(c). We focused on the prediction of the IP via the highest occupied KS eigenvalue, $\eps_\textrm{e-ho}$. 

We found that implementing the ensemble approach improves, on average, the correspondence of $\eps_\mathrm{ho}$ with the experimental IP for all xc functionals considered, changing the general tendency in the IP prediction from a gross underestimation to a smaller overestimation. 

For functionals that include a parameter, namely the hybrids PBEh(a) and ISOcc(c), we observed a rather weak dependence of $\eps_\textrm{e-ho}$ on the respective functional parameter, while yielding a roughly constant overestimation to the IP, with respect to experiment. This eases the so-called `parameter dilemma': there are no two optimal values of the functional's parameter originating from fitting to total energy-related quantities as opposed to fitting potential-related quantities. Instead, the parameter can be determined relying on energetics only, because of its weak influence on the value of $\eps_\textrm{e-ho}$. Indeed, the average relative error in the ionization potential, $\delta_\textrm{IP}$, equals approximately 15 \% for all ensemble-generalized xc functionals, as can be seen from Fig.~\ref{fig.varfuncs}. Surprisingly, such features of the underlying xc functional, as being local (LSDA), semi-local (PBE, BLYP) or non-local (B3LYP, PBEh, ISOcc), relying on features of the homogeneous electron gas (LSDA, PBE, PBEh, ISOcc) or not (BLYP, B3LYP), are of little relevance with respect to the IP prediction, once the functional is used in the ensemble-generalized form. 
We therefore conclude that upon ensemble generalization (Eq.~(\ref{eq.ET.gen})) all the functionals we tested share the same deficiency. It is most probably related to the remaining concavity of the $E(N)$ curve, due to the implicit dependence of the KS orbitals on $\alpha$. Therefore, future improvement in the IP prediction via $\eps_\mathrm{e-ho}$ may be achieved via formulating a correction that will remove the remaining concavity in $E(N)$.

\begin{acknowledgments} 

Financial support by the German-Israeli Foundation, the European Research Council and the Lise Meitner center for computational chemistry is gratefully acknowledged. E.K.\ is a recipient of the Levzion scholarship. T.S.\ acknowledges support from the Elite Network of Bavaria (``Macromolecular Science'' program).

\end{acknowledgments}

%\bibliography{library,bibliography_EK} 
%\bibliographystyle{aip} 

\end{document}

\title{Supplemental Material to: \emph{``Effect of ensemble generalization on the highest-occupied Kohn-Sham eigenvalue''}}

\author{Eli Kraisler}
\thanks{These authors contributed equally}
\affiliation{Department of Materials and Interfaces, Weizmann Institute of Science, Rehovoth 76100,Israel}

\author{Tobias Schmidt}
\thanks{These authors contributed equally}
\affiliation{Theoretical Physics IV, University of Bayreuth, 95440 Bayreuth, Germany}

\author{Stephan K\"{u}mmel}
\affiliation{Theoretical Physics IV, University of Bayreuth, 95440 Bayreuth, Germany}

\author{Leeor Kronik}
\affiliation{Department of Materials and Interfaces, Weizmann Institute of Science, Rehovoth 76100,Israel}

\date{\today}% It is always \today, today, but any date may be explicitly specified

\maketitle

In the following we present in tabular form the numerical data used to generate %ek3: old: a specific listing of the data that sets the foundation for 
Figs.\ (2)-(5) of the main text. %ek3:old: \red{article 'Effect of ensemble generalization on the highest-occupied Kohn-Sham energy level in atoms and diatomic molecules'}. 
For each system in the test set, we provide the negative of the unshifted highest occupied eigenvalue, $-\eps_{ho}$, as well as its shifted counterpart, $-\eps_{e-ho}$, for a variety of functionals discussed in the %ek3: old: main paper.
article. Table~\ref{table.var_funcs} contains results obtained with the functionals LSDA, BLYP, B3LYP and pure EXX, tables~\ref{table.pbeh} and~\ref{table.isocc} provide values obtained with the global hybrid functional PBEh($a$) and the local hybrid functional ISOcc($c$), as function of the parameter $a$ or $c$, respectively. Additionally, if calculated, the ionization potential evaluated by the $\Delta$SCF method is included in the tables. Finally, for each functional %put to task 
the averaged relative errors $\delta_{IP}$, evaluated with $-\eps_{ho}$, $-\eps_{e-ho}$ and $\Delta$SCF, together with the corresponding sign functions $S$, are presented. Our calculations were based on bond lengths and vertical ionization potentials determined by experiment as given in Ref.~\cite{HandChemPhys92} and \texttt{http://webbook.nist.gov}; the actual values are included in the tables for comparison.

%\section{Supplementary Material}\label{sec.suppl}%Writer: Tobi.
%--------------------------------------------------------------------------------------
%First Table: LDA, BLYP, B3LYP, EXX
%\begin{widetext}
\begin{longtable}{@{\extracolsep{\fill}}lcllllll}
%\caption{width=.75\textwidth}
\caption[width=40cm]{Unshifted and shifted highest occupied eigenvalue $-\eps_{ho}$ and $-\eps_{e-ho}$ for the system set evaluated with LSDA, BLYP, B3LYP and pure EXX. Additionally, for LSDA the IP computed via $\Delta$SCF is listed. All energy values are in \hartree.}
\label{table.var_funcs}\\\hline\hline
\multicolumn{1}{c}{}&\multicolumn{1}{c}{}&\multicolumn{1}{c}{}&\multicolumn{1}{l}{}& \multicolumn{4}{c}{Functional}\\
\multicolumn{1}{c}{System}&\multicolumn{1}{c}{$R_{AB}^{exp}$ (Bohr)}&\multicolumn{1}{c}{$IP_{exp}$}&\multicolumn{1}{l}{$IP_{DFT}$ via ...}& \multicolumn{1}{c}{LSDA} & \multicolumn{1}{c}{BLYP} & \multicolumn{1}{c}{B3LYP} & \multicolumn{1}{c}{EXX}\\ \hline
\endfirsthead
\multicolumn{8}{c}{{\bfseries \tablename\ \thetable{} -- continued from previous page}} \\
\hline \hline
\multicolumn{1}{c}{}&\multicolumn{1}{c}{}&\multicolumn{1}{c}{}&\multicolumn{1}{l}{}& \multicolumn{4}{c}{Functional}\\
\multicolumn{1}{c}{System}&\multicolumn{1}{c}{$R_{AB}^{exp}$ (Bohr)}&\multicolumn{1}{c}{$IP_{exp}$}&\multicolumn{1}{l}{$IP_{DFT}$ via ...}& \multicolumn{1}{c}{LSDA} & \multicolumn{1}{c}{BLYP} & \multicolumn{1}{c}{B3LYP} & \multicolumn{1}{c}{EXX}\\ \hline
\endhead

\hline \multicolumn{8}{c}{{Continued on next page}} \\ \hline
\endfoot

\hline \hline
\endlastfoot

H$_2$ & 1.4011 & 0.5669 & $-\eps_{ho}$   & 0.3772 & 0.3819 & 0.4313 & 0.5945 \\
      &        &        & $-\eps_{e-ho}$ & 0.6257 & 0.6329 & 0.6353 & 0.5945 \\ 
      &        &        & $\Delta$SCF & 0.5963 &        &        &        \\ 
LiH   & 3.0139 & 0.285  & $-\eps_{ho}$   & 0.1613 & 0.1588 & 0.1922 & 0.3011 \\ 
      &        &        & $-\eps_{e-ho}$ & 0.3424 & 0.3435 & 0.3437 & 0.3011 \\ 
      &        &        & $\Delta$SCF & 0.3017 &        &        &        \\ 
Li$_2$& 5.0518 & 0.1879 & $-\eps_{ho}$   & 0.1189 & 0.1127 & 0.1313 & 0.1812 \\ 
      &        &        & $-\eps_{e-ho}$ & 0.2147 & 0.2115 & 0.2123 & 0.1812 \\ 
      &        &        & $\Delta$SCF & 0.1954 &        &        &        \\ 
LiF   & 2.9553 & 0.4155 & $-\eps_{ho}$   & 0.2333 & 0.2254 & 0.2809 & 0.4760 \\ 
      &        &        & $-\eps_{e-ho}$   & 0.5835 & 0.5733 & 0.5622 & 0.4760 \\ 
      &        &        & $\Delta$SCF & 0.4542 &        &        &        \\ 
BeH   & 2.5368 & 0.3015 & $-\eps_{ho}$   & 0.1692 & 0.1666 & 0.1983 & 0.3096 \\ 
      &        &        & $-\eps_{e-ho}$ & 0.3152 & 0.3175 & 0.3193 & 0.3096 \\ 
      &        &        & $\Delta$SCF & 0.3057 &        &        &        \\ 
BH    & 2.3290 & 0.359  & $-\eps_{ho}$   & 0.2031 & 0.2011 & 0.2370 & 0.3461 \\ 
      &        &        & $-\eps_{e-ho}$ & 0.3876 & 0.3858 & 0.3869 & 0.3461 \\ 
      &        &        & $\Delta$SCF & 0.3558 &        &        &        \\ 
BO    & 2.2766 & 0.489  & $-\eps_{ho}$   & 0.3045 & 0.2462 & 0.2857 & 0.5199 \\ 
      &        &        & $-\eps_{e-ho}$ & 0.5064 & 0.4395 & 0.4429 & 0.5199 \\ 
      &        &        & $\Delta$SCF & 0.4765 &        &        &        \\ 
BF    & 2.3861 & 0.4087 & $-\eps_{ho}$   & 0.2508 & 0.3004 & 0.3491 & 0.4053 \\ 
      &        &        & $-\eps_{e-ho}$ & 0.4435 & 0.5038 & 0.5129 & 0.4053 \\ 
      &        &        & $\Delta$SCF & 0.4059 &        &        &        \\ 
CH    & 2.1163 & 0.391  & $-\eps_{ho}$   & 0.2185 & 0.2091 & 0.2562 & 0.4162 \\ 
      &        &        & $-\eps_{e-ho}$ & 0.4520 & 0.4411 & 0.4432 & 0.4162 \\ 
      &        &        & $\Delta$SCF & 0.4057 &        &        &        \\ 
CN    & 2.2144 & 0.4997 & $-\eps_{ho}$   & 0.3508 & 0.3404 & 0.3836 & 0.5316 \\ 
      &        &        & $-\eps_{e-ho}$ & 0.5932 & 0.5822 & 0.5778 & 0.5316 \\ 
      &        &        & $\Delta$SCF & 0.5376 &        &        &        \\ 
CO    & 2.1322 & 0.515  & $-\eps_{ho}$   & 0.3350 & 0.3311 & 0.3839 & 0.5526 \\ 
      &        &        & $-\eps_{e-ho}$ & 0.5769 & 0.5705 & 0.5780 & 0.5526 \\ 
      &        &        & $\Delta$SCF & 0.5176 &        &        &        \\ 
NH    & 1.9600 & 0.4958 & $-\eps_{ho}$   & 0.2928 & 0.2842 & 0.3404 & 0.5011 \\ 
      &        &        & $-\eps_{e-ho}$ & 0.5695$^{\dagger}$ & 0.5774$^{\dagger}$ & 0.5731$^{\dagger}$ & 0.5011 \\ 
      &        &        & $\Delta$SCF & 0.5066 &        &        &        \\ 
N$_2$ & 2.0743 & 0.5733 & $-\eps_{ho}$   & 0.3825 & 0.3770 & 0.4364 & 0.6303 \\ 
      &        &        & $-\eps_{e-ho}$ & 0.6251 & 0.6168 & 0.6294 & 0.6303 \\ 
      &        &        & $\Delta$SCF & 0.5740 &        &        &        \\ 
NO    & 2.1746 & 0.3405 & $-\eps_{ho}$   & 0.1690 & 0.1622 & 0.2206 & 0.4173 \\ 
      &        &        & $-\eps_{e-ho}$ & 0.4206 & 0.4116 & 0.4217 & 0.4173 \\ 
      &        &        & $\Delta$SCF & 0.3685 &        &        &        \\ 
OH    & 1.8324 & 0.4784 & $-\eps_{ho}$   & 0.2740 & 0.2629 & 0.3203 & 0.4984 \\ 
      &        &        & $-\eps_{e-ho}$ & 0.5879 & 0.5740 & 0.5715 & 0.4984 \\ 
      &        &        & $\Delta$SCF & 0.4957 &        &        &        \\ 
O$_2$ & 2.2819 & 0.4531 & $-\eps_{ho}$   & 0.2547 & 0.2502 & 0.3173 & 0.5571 \\ 
      &        &        & $-\eps_{e-ho}$ & 0.5316 & 0.5265 & 0.5396 & 0.5571 \\ 
      &        &        & $\Delta$SCF & 0.4683 &        &        &        \\ 
FH    & 1.7326 & 0.5924 & $-\eps_{ho}$   & 0.3608 & 0.3544 & 0.4203 & 0.6453 \\ 
      &        &        & $-\eps_{e-ho}$ & 0.7391 & 0.7313 & 0.7235 & 0.6453 \\ 
      &        &        & $\Delta$SCF & 0.6172 &        &        &        \\ 
F$_2$ & 2.6695 & 0.5769 & $-\eps_{ho}$   & 0.3544 & 0.3481 & 0.4205 & 0.6667 \\ 
      &        &        & $-\eps_{e-ho}$ & 0.6478 & 0.6406 & 0.6552 & 0.6667 \\ 
      &        &        & $\Delta$SCF & 0.5744 &        &        &        \\ 
H     &        & 0.4997 & $-\eps_{ho}$   & 0.2690 & 0.2722 & 0.3191 & 0.5000 \\ 
      &        &        & $-\eps_{e-ho}$ & 0.4787 & 0.4979 & 0.4991 & 0.5000 \\ 
      &        &        & $\Delta$SCF & 0.4787 &        &        &        \\ 
Li    &        & 0.1981 & $-\eps_{ho}$   & 0.1163 & 0.1114 & 0.1311 & 0.1962 \\ 
      &        &        & $-\eps_{e-ho}$ & 0.2013 & 0.2034 & 0.2041 & 0.1962 \\ 
      &        &        & $\Delta$SCF & 0.2011 &        &        &        \\ 
Be    &        & 0.3426 & $-\eps_{ho}$   & 0.2057 & 0.2009 & 0.2290 & 0.3089 \\ 
      &        &        & $-\eps_{e-ho}$ & 0.3447 & 0.3439 & 0.3454 & 0.3089 \\ 
      &        &        & $\Delta$SCF & 0.3318 &        &        &        \\ 
B     &        & 0.3049 & $-\eps_{ho}$   & 0.1509 & 0.1494 & 0.1866 & 0.3170 \\ 
      &        &        & $-\eps_{e-ho}$ & 0.3380 & 0.3382 & 0.3390 & 0.3170 \\ 
      &        &        & $\Delta$SCF & 0.3175 &        &        &        \\ 
C     &        & 0.4138 & $-\eps_{ho}$   & 0.2249 & 0.2182 & 0.2662 & 0.4378 \\ 
      &        &        & $-\eps_{e-ho}$ & 0.4714 & 0.4667 & 0.4661 & 0.4378 \\ 
      &        &        & $\Delta$SCF & 0.4313 &        &        &        \\ 
N     &        & 0.5341 & $-\eps_{ho}$   & 0.3085 & 0.2970 & 0.3560 & 0.5705 \\ 
      &        &        & $-\eps_{e-ho}$ & 0.6115 & 0.5996 & 0.5957 & 0.5705 \\ 
      &        &        & $\Delta$SCF & 0.5512 &        &        &        \\ 
O     &        & 0.5005 & $-\eps_{ho}$   & 0.2737 & 0.2803 & 0.3366 & 0.5193 \\ 
      &        &        & $-\eps_{e-ho}$ & 0.5962 & 0.6045 & 0.5984 & 0.5193 \\ 
      &        &        & $\Delta$SCF & 0.5146 &        &        &        \\ 
F     &        & 0.6403 & $-\eps_{ho}$   & 0.3808 & 0.3797 & 0.4478 & 0.6779 \\ 
      &        &        & $-\eps_{e-ho}$ & 0.7679 & 0.7677 & 0.7598 & 0.6779 \\ 
      &        &        & $\Delta$SCF & 0.6598 &        &        &        \\ \hline
\multicolumn{1}{c}{}&\multicolumn{1}{c}{}&\multicolumn{1}{c}{}&\multicolumn{1}{l}{$\delta_{IP}$ (\%) via $-\eps_{ho}$}& \multicolumn{1}{r}{\textbf{41.45}} & \multicolumn{1}{r}{\textbf{42.62}} & \multicolumn{1}{r}{\textbf{31.50}} &\multicolumn{1}{r}{\textbf{9.19}}\\

\multicolumn{1}{c}{}&\multicolumn{1}{c}{}&\multicolumn{1}{c}{}&\multicolumn{1}{l}{$\delta_{IP}$ (\%) via $-\eps_{e-ho}$}& \multicolumn{1}{r}{\textbf{16.42}} & \multicolumn{1}{r}{\textbf{15.48}} & \multicolumn{1}{r}{\textbf{15.48}} &\multicolumn{1}{r}{\textbf{9.19}}\\

\multicolumn{1}{c}{}&\multicolumn{1}{c}{}&\multicolumn{1}{c}{}&\multicolumn{1}{l}{$\delta_{IP}$ (\%) via $\Delta$SCF }& \multicolumn{1}{r}{\textbf{4.17}} &  &  &\\

\multicolumn{1}{c}{}&\multicolumn{1}{c}{}&\multicolumn{1}{c}{}&\multicolumn{1}{l}{$S$ via $-\eps_{ho}\ \ \ $ }& \multicolumn{1}{r}{\textbf{-1.00}} & \multicolumn{1}{r}{\textbf{-1.00}} & \multicolumn{1}{r}{\textbf{-1.00}} &\multicolumn{1}{r}{\textbf{0.62}}\\

\multicolumn{1}{c}{}&\multicolumn{1}{c}{}&\multicolumn{1}{c}{}&\multicolumn{1}{l}{$S$ via $-\eps_{e-ho}$ }& \multicolumn{1}{r}{\textbf{0.92}} & \multicolumn{1}{r}{\textbf{0.92}} & \multicolumn{1}{r}{\textbf{0.92}} &\multicolumn{1}{r}{\textbf{0.62}}\\
%\end{tabular*}
\end{longtable}\vspace{-0.4cm}

\begin{description}
\item[$^\dagger$]\footnotesize In this case, the states marking the highest occupied eigenvalue before and after the shift are not the same, but they belong to different spin channels.
\end{description}
%\end{widetext}
\newpage

%--------------------------------------------------------------------------------------
%Second Table: PBEh
\begin{landscape}
\setlength{\LTcapwidth}{24cm}
\begin{longtable}{@{\extracolsep{\fill}}lclllllllllllllll}
\caption{Unshifted and shifted highest occupied eigenvalue $-\eps_{ho}$ and $-\eps_{e-ho}$ for the system set evaluated with the global hybrid functional PBEh($a$) in dependence on the functional parameter $a$. Additionally, for pure PBE (i.e., $a=0$) the IP computed via $\Delta$SCF is listed. All energy values are in \hartree.}\label{table.pbeh}\\\hline\hline
\multicolumn{1}{c}{}&\multicolumn{1}{c}{}&\multicolumn{1}{c}{}&\multicolumn{1}{l}{}& \multicolumn{13}{c}{Functional parameter $a$}\\
\multicolumn{1}{c}{System}&\multicolumn{1}{c}{$R_{AB}^{exp}$ (Bohr)}&\multicolumn{1}{c}{$IP_{exp}$}&\multicolumn{1}{l}{$IP_{DFT}$ via ...}& \multicolumn{1}{c}{0} & \multicolumn{1}{c}{0.1} & \multicolumn{1}{c}{0.2}&\multicolumn{1}{c}{0.25}&\multicolumn{1}{c}{0.3}& \multicolumn{1}{c}{0.4}& \multicolumn{1}{c}{0.5}& \multicolumn{1}{c}{0.6}& \multicolumn{1}{c}{0.7}& \multicolumn{1}{c}{0.75}& \multicolumn{1}{c}{0.8}& \multicolumn{1}{c}{0.9}& \multicolumn{1}{c}{1.0}\\ \hline
\endfirsthead
\multicolumn{17}{c}{{\bfseries \tablename\ \thetable{} -- continued from previous page}} \\
\hline \hline
\multicolumn{1}{c}{}&\multicolumn{1}{c}{}&\multicolumn{1}{c}{}&\multicolumn{1}{l}{}& \multicolumn{13}{c}{Functional parameter $a$}\\
\multicolumn{1}{c}{System}&\multicolumn{1}{c}{$R_{AB}^{exp}$ (Bohr)}&\multicolumn{1}{c}{$IP_{exp}$}&\multicolumn{1}{l}{$IP_{DFT}$ via ...}& \multicolumn{1}{c}{0} & \multicolumn{1}{c}{0.1} & \multicolumn{1}{c}{0.2}&\multicolumn{1}{c}{0.25}&\multicolumn{1}{c}{0.3}& \multicolumn{1}{c}{0.4}& \multicolumn{1}{c}{0.5}& \multicolumn{1}{c}{0.6}& \multicolumn{1}{c}{0.7}& \multicolumn{1}{c}{0.75}&\multicolumn{1}{c}{0.8}& \multicolumn{1}{c}{0.9}& \multicolumn{1}{c}{1.0}\\ \hline
\endhead

\hline \multicolumn{17}{c}{{Continued on next page}} \\ \hline
\endfoot

\hline \hline
\endlastfoot

H$_2$  & 1.4011 & 0.5669 &  $-\eps_{ho}$  & 0.3815 & 0.4053 & 0.4292 & 0.4412 & 0.4532 & 0.4772 & 0.5013 & 0.5256 & 0.5498 & 0.5620 & 0.5742 & 0.5986 & 0.6230 \\ 
       &          &          &  $-\eps_{e-ho}$ & 0.6268 & 0.6269 & 0.6271 & 0.6272 & 0.6273 & 0.6275 & 0.6278 & 0.6281 & 0.6284 & 0.6285 & 0.6287 & 0.6290 & 0.6294 \\ 
       &          &          &  $\Delta$SCF & 0.5965 &  &  &  &  &  &  &  &  &  &  &  &\\ 
LiH    & 3.0139 & 0.285 &  $-\eps_{ho}$  & 0.1603 & 0.1759 & 0.1917 & 0.1998 & 0.2079 & 0.2242 & 0.2407 & 0.2575 & 0.2744 & 0.2830 & 0.2916 & 0.3088 & 0.3263 \\ 
       &          &          &  $-\eps_{e-ho}$ & 0.3393 & 0.3382 & 0.3372 & 0.3367 & 0.3363 & 0.3355 & 0.3347 & 0.3340 & 0.3333 & 0.3330 & 0.3327 & 0.3321 & 0.3316 \\ 
       &          &          &  $\Delta$SCF & 0.2959 &  &  &  &  &  &  &  &  &  &  &  &  \\ 
Li$_2$ & 5.0518 & 0.1879 &  $-\eps_{ho}$  & 0.1185 & 0.1268 & 0.1351 & 0.1392 & 0.1434 & 0.1518 & 0.1602 & 0.1687 & 0.1772 & 0.1814 & 0.1857 & 0.1943 & 0.2029 \\ 
       &          &          &  $-\eps_{e-ho}$ & 0.2132 & 0.2124 & 0.2117 & 0.2114 & 0.2110 & 0.2104 & 0.2097 & 0.2091 & 0.2085 & 0.2082 & 0.2079 & 0.2073 & 0.2068 \\ 
       &          &          &  $\Delta$SCF & 0.1931 &  &  &  &  &  &  &  &  &  &  &  &  \\ 
LiF    & 2.9553 & 0.4155 &  $-\eps_{ho}$  & 0.2252 & 0.2515 & 0.2783 & 0.2920 & 0.3058 & 0.3338 & 0.3622 & 0.3912 & 0.4206 & 0.4355 & 0.4504 & 0.4806 & 0.5112 \\ 
       &          &          &  $-\eps_{e-ho}$  & 0.5731 & 0.5660 & 0.5592 & 0.5559 & 0.5527 & 0.5464 & 0.5403 & 0.5344 & 0.5287 & 0.5259 & 0.5232 & 0.5179 & 0.5127 \\ 
       &          &          &  $\Delta$SCF & 0.4437 &  &  &  &  &  &  &  &  &  &  &  &  \\ 
BeH    & 2.5368 & 0.3015 &  $-\eps_{ho}$  & 0.1728 & 0.1876 & 0.2024 & 0.2098 & 0.2173 & 0.2322 & 0.2472 & 0.2622 & 0.2773 & 0.2848 & 0.2924 & 0.3075 & 0.3227 \\ 
       &          &          &  $-\eps_{e-ho}$ & 0.3197 & 0.3198 & 0.3199 & 0.3199 & 0.3200 & 0.3201 & 0.3202 & 0.3203 & 0.3204 & 0.3205 & 0.3205 & 0.3206 & 0.3207 \\ 
       &          &          &  $\Delta$SCF & 0.3099 &  &  &  &  &  &  &  &  &  &  &  &  \\ 
BH     & 2.3290 & 0.359 &  $-\eps_{ho}$  & 0.2033 & 0.2201 & 0.2370 & 0.2455 & 0.2540 & 0.2710 & 0.2881 & 0.3052 & 0.3224 & 0.3310 & 0.3396 & 0.3569 & 0.3742 \\ 
       &          &          &  $-\eps_{e-ho}$ & 0.3846 & 0.3839 & 0.3833 & 0.3829 & 0.3826 & 0.3820 & 0.3813 & 0.3807 & 0.3801 & 0.3798 & 0.3796 & 0.3790 & 0.3785 \\ 
       &          &          &  $\Delta$SCF & 0.3520 &  &  &  &  &  &  &  &  &  &  &  &  \\ 
BO     & 2.2766 & 0.489 &  $-\eps_{ho}$  & 0.2492 & 0.2678 & 0.2864 & 0.2957 & 0.3051 & 0.3238 & 0.3424 & 0.3611 & 0.3798 & 0.3892 & 0.3986 & 0.4173 & 0.4361 \\ 
       &          &          &  $-\eps_{e-ho}$ & 0.4382 & 0.4386 & 0.4389 & 0.4391 & 0.4392 & 0.4395 & 0.4398 & 0.4400 & 0.4402 & 0.4404 & 0.4404 & 0.4406 & 0.4408 \\ 
       &          &          &  $\Delta$SCF & 0.3997 &  &  &  &  &  &  &  &  &  &  &  &  \\ 
BF     & 2.3861 & 0.4087 &  $-\eps_{ho}$  & 0.3063 & 0.3296 & 0.3529 & 0.3646 & 0.3762 & 0.3995 & 0.4227 & 0.4460 & 0.4692 & 0.4808 & 0.4924 & 0.5156 & 0.5387 \\ 
       &          &          &  $-\eps_{e-ho}$ & 0.5063 & 0.5100 & 0.5135 & 0.5153 & 0.5170 & 0.5203 & 0.5235 & 0.5266 & 0.5296 & 0.5311 & 0.5325 & 0.5353 & 0.5380 \\ 
       &          &          &  $\Delta$SCF & 0.4770 &  &  &  &  &  &  &  &  &  &  &  &  \\ 
CH     & 2.1163 & 0.391 &  $-\eps_{ho}$  & 0.2155 & 0.2378 & 0.2602 & 0.2715 & 0.2828 & 0.3056 & 0.3284 & 0.3514 & 0.3745 & 0.3860 & 0.3976 & 0.4208 & 0.4442 \\ 
       &          &          &  $-\eps_{e-ho}$ & 0.4465 & 0.4462 & 0.4460 & 0.4459 & 0.4459 & 0.4457 & 0.4456 & 0.4455 & 0.4455 & 0.4454 & 0.4454 & 0.4454 & 0.4454 \\ 
       &          &          &  $\Delta$SCF & 0.4005 &  &  &  &  &  &  &  &  &  &  &  &  \\ 
CN     & 2.2144 & 0.4997 &  $-\eps_{ho}$  & 0.3452 & 0.3655 & 0.3859 & 0.3961 & 0.4064 & 0.4271 & 0.4480 & 0.4694 & 0.4912 & 0.5023 & 0.5136 & 0.5368 & 0.5608 \\ 
       &          &          &  $-\eps_{e-ho}$ & 0.5868 & 0.5832 & 0.5798 & 0.5781 & 0.5765 & 0.5734 & 0.5705 & 0.5680 & 0.5659 & 0.5650 & 0.5642 & 0.5629 & 0.5621 \\ 
       &          &          &  $\Delta$SCF & 0.5305 &  &  &  &  &  &  &  &  &  &  &  &  \\ 
CO     & 2.1322 & 0.515 &  $-\eps_{ho}$  & 0.3322 & 0.3574 & 0.3827 & 0.3953 & 0.4080 & 0.4333 & 0.4587 & 0.4841 & 0.5095 & 0.5222 & 0.5349 & 0.5604 & 0.5859 \\ 
       &          &          &  $-\eps_{e-ho}$ & 0.5685 & 0.5709 & 0.5732 & 0.5744 & 0.5755 & 0.5777 & 0.5799 & 0.5821 & 0.5842 & 0.5852 & 0.5862 & 0.5882 & 0.5902 \\ 
       &          &          &  $\Delta$SCF & 0.5092 &  &  &  &  &  &  &  &  &  &  &  &  \\ 
NH     & 1.9600 & 0.4958 &  $-\eps_{ho}$  & 0.2911 & 0.3179 & 0.3449 & 0.3584 & 0.3720 & 0.3994 & 0.4270 & 0.4518 & 0.4744 & 0.4858 & 0.4971 & 0.5199 & 0.5428 \\ 
       &          &          &  $-\eps_{e-ho}$ & 0.5742$^{\dagger}$ & 0.5712$^{\dagger}$ & 0.5683$^{\dagger}$ & 0.5669$^{\dagger}$ & 0.5656$^{\dagger}$ & 0.5629$^{\dagger}$ & 0.5602$^{\dagger}$ & 0.5576 & 0.5551 & 0.5539 & 0.5528 & 0.5504 & 0.5481 \\ 
       &          &          &  $\Delta$SCF & 0.5079 &  &  &  &  &  &  &  &  &  &  &  &  \\ 
N$_2$  & 2.0743 & 0.5733 &  $-\eps_{ho}$  & 0.3773 & 0.4058 & 0.4343 & 0.4486 & 0.4629 & 0.4916 & 0.5204 & 0.5491 & 0.5780 & 0.5924 & 0.6068 & 0.6357 & 0.6647 \\ 
       &          &          &  $-\eps_{e-ho}$ & 0.6172 & 0.6220 & 0.6269 & 0.6293 & 0.6318 & 0.6367 & 0.6417 & 0.6467 & 0.6517 & 0.6542 & 0.6568 & 0.6618 & 0.6669 \\ 
       &          &          &  $\Delta$SCF & 0.5655 &  &  &  &  &  &  &  &  &  &  &  &  \\ 
NO     & 2.1746 & 0.3405 &  $-\eps_{ho}$  & 0.1635 & 0.1914 & 0.2195 & 0.2337 & 0.2479 & 0.2764 & 0.3050 & 0.3338 & 0.3627 & 0.3771 & 0.3916 & 0.4207 & 0.4498 \\ 
       &          &          &  $-\eps_{e-ho}$ & 0.4125 & 0.4162 & 0.4200 & 0.4219 & 0.4238 & 0.4276 & 0.4314 & 0.4353 & 0.4392 & 0.4411 & 0.4431 & 0.4469 & 0.4508 \\ 
       &          &          &  $\Delta$SCF & 0.3614 &  &  &  &  &  &  &  &  &  &  &  &  \\ 
OH     & 1.8324 & 0.4784 &  $-\eps_{ho}$  & 0.2636 & 0.2908 & 0.3182 & 0.3320 & 0.3458 & 0.3736 & 0.4016 & 0.4297 & 0.4579 & 0.4721 & 0.4863 & 0.5148 & 0.5435 \\ 
       &          &          &  $-\eps_{e-ho}$ & 0.5727 & 0.5697 & 0.5668 & 0.5654 & 0.5639 & 0.5612 & 0.5585 & 0.5558 & 0.5533 & 0.5520 & 0.5508 & 0.5483 & 0.5460 \\ 
       &          &          &  $\Delta$SCF & 0.4804 &  &  &  &  &  &  &  &  &  &  &  &  \\ 
O$_2$  & 2.2819 & 0.4531 &  $-\eps_{ho}$  & 0.2507 & 0.2829 & 0.3155 & 0.3319 & 0.3484 & 0.3815 & 0.4150 & 0.4486 & 0.4825 & 0.4995 & 0.5166 & 0.5509 & 0.5854 \\ 
       &          &          &  $-\eps_{e-ho}$ & 0.5261 & 0.5315 & 0.5370 & 0.5398 & 0.5427 & 0.5485 & 0.5545 & 0.5606 & 0.5668 & 0.5699 & 0.5731 & 0.5796 & 0.5862 \\ 
       &          &          &  $\Delta$SCF & 0.4635 &  &  &  &  &  &  &  &  &  &  &  &  \\ 
FH     & 1.7326 & 0.5924 &  $-\eps_{ho}$  & 0.3547 & 0.3864 & 0.4183 & 0.4344 & 0.4505 & 0.4830 & 0.5157 & 0.5487 & 0.5818 & 0.5985 & 0.6152 & 0.6487 & 0.6825 \\ 
       &          &          &  $-\eps_{e-ho}$ & 0.7312 & 0.7260 & 0.7209 & 0.7183 & 0.7158 & 0.7110 & 0.7062 & 0.7016 & 0.6971 & 0.6948 & 0.6926 & 0.6883 & 0.6841 \\ 
       &          &          &  $\Delta$SCF & 0.6093 &  &  &  &  &  &  &  &  &  &  &  &  \\ 
F$_2$  & 2.6695 & 0.5769 &  $-\eps_{ho}$  & 0.3475 & 0.3824 & 0.4174 & 0.4351 & 0.4528 & 0.4883 & 0.5240 & 0.5599 & 0.5960 & 0.6142 & 0.6323 & 0.6688 & 0.7054 \\ 
       &          &          &  $-\eps_{e-ho}$ & 0.6398 & 0.6457 & 0.6518 & 0.6549 & 0.6580 & 0.6644 & 0.6711 & 0.6778 & 0.6847 & 0.6882 & 0.6917 & 0.6989 & 0.7062 \\ 
       &          &          &  $\Delta$SCF & 0.5669 &  &  &  &  &  &  &  &  &  &  &  &  \\ 
H      &          & 0.4997 &  $-\eps_{ho}$  & 0.2791 & 0.3016 & 0.3242 & 0.3355 & 0.3469 & 0.3698 & 0.3928 & 0.4159 & 0.4392 & 0.4508 & 0.4625 & 0.4860 & 0.5096 \\ 
       &          &          &  $-\eps_{e-ho}$ & 0.5000 & 0.5005 & 0.5010 & 0.5013 & 0.5016 & 0.5022 & 0.5028 & 0.5034 & 0.5040 & 0.5043 & 0.5047 & 0.5053 & 0.5060 \\ 
       &          &          &  $\Delta$SCF & 0.5000 &  &  &  &  &  &  &  &  &  &  &  &  \\ 
Li     &          & 0.1981 &  $-\eps_{ho}$   & 0.1186 & 0.1275 & 0.1363 & 0.1407 & 0.1451 & 0.1540 & 0.1628 & 0.1716 & 0.1803 & 0.1847 & 0.1891 & 0.1979 & 0.2066 \\ 
       &          &          &  $-\eps_{e-ho}$ & 0.2055 & 0.2052 & 0.2049 & 0.2048 & 0.2047 & 0.2044 & 0.2041 & 0.2039 & 0.2036 & 0.2035 & 0.2034 & 0.2032 & 0.2029 \\ 
       &          &          &  $\Delta$SCF  & 0.2053 &  &  &  &  &  &  &  &  &  &  &  &  \\ 
Be     &          & 0.3426 &  $-\eps_{ho}$   & 0.2061 & 0.2190 & 0.2320 & 0.2384 & 0.2449 & 0.2578 & 0.2708 & 0.2837 & 0.2966 & 0.3031 & 0.3095 & 0.3224 & 0.3353 \\ 
       &          &          &  $-\eps_{e-ho}$ & 0.3436 & 0.3431 & 0.3427 & 0.3425 & 0.3423 & 0.3419 & 0.3414 & 0.3410 & 0.3406 & 0.3404 & 0.3402 & 0.3399 & 0.3395 \\ 
       &          &          &  $\Delta$SCF  & 0.3307 &  &  &  &  &  &  &  &  &  &  &  &  \\ 
B      &          & 0.3049 &  $-\eps_{ho}$   & 0.1534 & 0.1708 & 0.1884 & 0.1973 & 0.2062 & 0.2241 & 0.2422 & 0.2603 & 0.2786 & 0.2878 & 0.2970 & 0.3156 & 0.3342 \\ 
       &          &          &  $-\eps_{e-ho}$ & 0.3392 & 0.3387 & 0.3383 & 0.3381 & 0.3378 & 0.3374 & 0.3371 & 0.3368 & 0.3365 & 0.3363 & 0.3362 & 0.3360 & 0.3358 \\ 
       &          &          &  $\Delta$SCF  & 0.3186 &  &  &  &  &  &  &  &  &  &  &  &  \\ 
C      &          & 0.4138 &  $-\eps_{ho}$   & 0.2241 & 0.2467 & 0.2695 & 0.2810 & 0.2925 & 0.3157 & 0.3390 & 0.3626 & 0.3862 & 0.3981 & 0.4101 & 0.4340 & 0.4581 \\ 
       &          &          &  $-\eps_{e-ho}$ & 0.4697 & 0.4683 & 0.4670 & 0.4663 & 0.4657 & 0.4645 & 0.4633 & 0.4622 & 0.4612 & 0.4607 & 0.4602 & 0.4592 & 0.4583 \\ 
       &          &          &  $\Delta$SCF  & 0.4308 &  &  &  &  &  &  &  &  &  &  &  &  \\ 
N      &          & 0.5341 &  $-\eps_{ho}$   & 0.3052 & 0.3330 & 0.3611 & 0.3752 & 0.3894 & 0.4179 & 0.4467 & 0.4756 & 0.5048 & 0.5195 & 0.5342 & 0.5637 & 0.5934 \\ 
       &          &          &  $-\eps_{e-ho}$ & 0.6058 & 0.6007 & 0.5999 & 0.5995 & 0.5992 & 0.5985 & 0.5979 & 0.5974 & 0.5970 & 0.5968 & 0.5966 & 0.5963 & 0.5961 \\ 
       &          &          &  $\Delta$SCF  & 0.5414 &  &  &  &  &  &  &  &  &  &  &  &  \\ 
O      &          & 0.5005 &  $-\eps_{ho}$   & 0.2794 & 0.3067 & 0.3343 & 0.3481 & 0.3621 & 0.3901 & 0.4183 & 0.4467 & 0.4752 & 0.4896 & 0.5040 & 0.5329 & 0.5619 \\ 
       &          &          &  $-\eps_{e-ho}$ & 0.6014 & 0.5974 & 0.5936 & 0.5917 & 0.5899 & 0.5862 & 0.5827 & 0.5792 & 0.5759 & 0.5743 & 0.5726 & 0.5695 & 0.5664 \\ 
       &          &          &  $\Delta$SCF  & 0.5167 &  &  &  &  &  &  &  &  &  &  &  &  \\ 
F      &          & 0.6403 &  $-\eps_{ho}$   & 0.3788 & 0.4116 & 0.4448 & 0.4614 & 0.4782 & 0.5118 & 0.5458 & 0.5799 & 0.6142 & 0.6315 & 0.6489 & 0.6837 & 0.7186 \\ 
       &          &          &  $-\eps_{e-ho}$ & 0.7663 & 0.7611 & 0.7561 & 0.7537 & 0.7512 & 0.7465 & 0.7419 & 0.7374 & 0.7329 & 0.7308 & 0.7287 & 0.7245 & 0.7204 \\ 
       &          &          &  $\Delta$SCF  & 0.6488 &  &  &  &  &  &  &  &  &  &  &  &  \\  \hline
\multicolumn{1}{c}{}&\multicolumn{1}{c}{}&\multicolumn{1}{c}{}&\multicolumn{1}{l}{$\delta_{IP}$ (\%) via $-\eps_{ho}$}& \multicolumn{1}{r}{\textbf{41.72}} & \multicolumn{1}{r}{\textbf{36.44}}& \multicolumn{1}{r}{\textbf{31.13}}& \multicolumn{1}{r}{\textbf{28.48}}& \multicolumn{1}{r}{\textbf{25.82}}& \multicolumn{1}{r}{\textbf{20.52}}& \multicolumn{1}{r}{\textbf{15.30}}& \multicolumn{1}{r}{\textbf{10.35}}& \multicolumn{1}{r}{\textbf{6.35}}& \multicolumn{1}{r}{\textbf{5.45}}& \multicolumn{1}{r}{\textbf{5.87}}& \multicolumn{1}{r}{\textbf{9.51}}& \multicolumn{1}{r}{\textbf{14.47}}\\

\multicolumn{1}{c}{}&\multicolumn{1}{c}{}&\multicolumn{1}{c}{}&\multicolumn{1}{l}{$\delta_{IP}$ (\%) via $-\eps_{e-ho}$}& \multicolumn{1}{r}{\textbf{15.48}} & \multicolumn{1}{r}{\textbf{15.20}}& \multicolumn{1}{r}{\textbf{14.99}}& \multicolumn{1}{r}{\textbf{14.90}}& \multicolumn{1}{r}{\textbf{14.83}}& \multicolumn{1}{r}{\textbf{14.71}}& \multicolumn{1}{r}{\textbf{14.63}}& \multicolumn{1}{r}{\textbf{14.60}}& \multicolumn{1}{r}{\textbf{14.61}}& \multicolumn{1}{r}{\textbf{14.63}}& \multicolumn{1}{r}{\textbf{14.66}}& \multicolumn{1}{r}{\textbf{14.75}}& \multicolumn{1}{r}{\textbf{14.88}}\\

\multicolumn{1}{c}{}&\multicolumn{1}{c}{}&\multicolumn{1}{c}{}&\multicolumn{1}{l}{$\delta_{IP}$ (\%) via $\Delta$SCF}& \multicolumn{1}{r}{\textbf{3.41}} & \multicolumn{1}{r}{\textbf{}}& \multicolumn{1}{r}{\textbf{}}& \multicolumn{1}{r}{\textbf{}}& \multicolumn{1}{r}{\textbf{}}& \multicolumn{1}{r}{\textbf{}}& \multicolumn{1}{r}{\textbf{}}& \multicolumn{1}{r}{\textbf{}}& \multicolumn{1}{r}{\textbf{}}& \multicolumn{1}{r}{\textbf{}}& \multicolumn{1}{r}{\textbf{}}& \multicolumn{1}{r}{\textbf{}}& \multicolumn{1}{r}{\textbf{}}\\

\multicolumn{1}{c}{}&\multicolumn{1}{c}{}&\multicolumn{1}{c}{}&\multicolumn{1}{l}{$S$ via $-\eps_{ho}$}& \multicolumn{1}{r}{\textbf{-1.00}} & \multicolumn{1}{r}{\textbf{-1.00}}& \multicolumn{1}{r}{\textbf{-1.00}}& \multicolumn{1}{r}{\textbf{-1.00}}& \multicolumn{1}{r}{\textbf{-1.00}}& \multicolumn{1}{r}{\textbf{-1.00}}& \multicolumn{1}{r}{\textbf{-1.00}}& \multicolumn{1}{r}{\textbf{-1.00}}& \multicolumn{1}{r}{\textbf{-0.62}}& \multicolumn{1}{r}{\textbf{-0.38}}& \multicolumn{1}{r}{\textbf{0.31}}& \multicolumn{1}{r}{\textbf{0.69}}& \multicolumn{1}{r}{\textbf{0.92}}\\

\multicolumn{1}{c}{}&\multicolumn{1}{c}{}&\multicolumn{1}{c}{}&\multicolumn{1}{l}{$S$ via $-\eps_{e-ho}$}& \multicolumn{1}{r}{\textbf{1.00}} & \multicolumn{1}{r}{\textbf{1.00}}& \multicolumn{1}{r}{\textbf{1.00}}& \multicolumn{1}{r}{\textbf{0.92}}& \multicolumn{1}{r}{\textbf{0.92}}& \multicolumn{1}{r}{\textbf{0.92}}& \multicolumn{1}{r}{\textbf{0.92}}& \multicolumn{1}{r}{\textbf{0.92}}& \multicolumn{1}{r}{\textbf{0.92}}& \multicolumn{1}{r}{\textbf{0.92}}& \multicolumn{1}{r}{\textbf{0.92}}& \multicolumn{1}{r}{\textbf{0.92}}& \multicolumn{1}{r}{\textbf{0.92}}\\
\end{longtable}\vspace{-0.4cm}
\begin{description}
\item[$^\dagger$]\footnotesize In this case, the states marking the highest occupied eigenvalue before and after the shift are not the same, but they belong to different spin-channels.
\end{description}
\newpage
\end{landscape}

%--------------------------------------------------------------------------------------
%Second Table: ISOcc
\begin{landscape}
\setlength{\LTcapwidth}{24cm}
\begin{longtable}{@{\extracolsep{\fill}}lclllllllllllllll}
\caption{Unshifted and shifted highest occupied eigenvalue $-\eps_{ho}$ and $-\eps_{e-ho}$ for the system set evaluated with the local hybrid functional ISOcc($c$) in dependence on the functional parameter $c$. Additionally, the IP computed via $\Delta$SCF is listed. All energy values are in \hartree.}\label{table.isocc}\\\hline\hline
\multicolumn{1}{c}{}&\multicolumn{1}{c}{}&\multicolumn{1}{c}{}&\multicolumn{1}{l}{}& \multicolumn{13}{c}{Functional parameter $c$}\\
\multicolumn{1}{c}{System}&\multicolumn{1}{c}{$R_{AB}^{exp}$ (Bohr)}&\multicolumn{1}{c}{$IP_{exp}$}&\multicolumn{1}{l}{$IP_{DFT}$ via ...}& \multicolumn{1}{c}{0} & \multicolumn{1}{c}{0.1} & \multicolumn{1}{c}{0.3}&\multicolumn{1}{c}{0.5}&\multicolumn{1}{c}{1.0}& \multicolumn{1}{c}{1.5}& \multicolumn{1}{c}{2.0}& \multicolumn{1}{c}{2.5}& \multicolumn{1}{c}{3.0}& \multicolumn{1}{c}{3.5}& \multicolumn{1}{c}{4.0}& \multicolumn{1}{c}{4.5}& \multicolumn{1}{c}{5.0}\\ \hline
\endfirsthead
\multicolumn{17}{c}{{\bfseries \tablename\ \thetable{} -- continued from previous page}} \\
\hline \hline
\multicolumn{1}{c}{}&\multicolumn{1}{c}{}&\multicolumn{1}{c}{}&\multicolumn{1}{l}{}& \multicolumn{13}{c}{Functional parameter $c$}\\
\multicolumn{1}{c}{System}&\multicolumn{1}{c}{$R_{AB}^{exp}$ (Bohr)}&\multicolumn{1}{c}{$IP_{exp}$}&\multicolumn{1}{l}{$IP_{DFT}$ via ...}& \multicolumn{1}{c}{0} & \multicolumn{1}{c}{0.1} & \multicolumn{1}{c}{0.3}&\multicolumn{1}{c}{0.5}&\multicolumn{1}{c}{1.0}& \multicolumn{1}{c}{1.5}& \multicolumn{1}{c}{2.0}& \multicolumn{1}{c}{2.5}& \multicolumn{1}{c}{3.0}& \multicolumn{1}{c}{3.5}& \multicolumn{1}{c}{4.0}& \multicolumn{1}{c}{4.5}& \multicolumn{1}{c}{5.0}\\ \hline
\endhead

\hline \multicolumn{17}{c}{{Continued on next page}} \\ \hline
\endfoot

\hline \hline
\endlastfoot

H$_2$  & 1.4011 & 0.5669 &  $-\eps_{ho}$  & 0.3772 & 0.3954 & 0.4249 & 0.4475 & 0.4860 & 0.5108 & 0.5282 & 0.5413 & 0.5515 & 0.5597 & 0.5665 & 0.5722 & 0.5771 \\ 
       &          &          &  $-\eps_{e-ho}$ & 0.6036 & 0.6122 & 0.6246 & 0.6332 & 0.6465 & 0.6543 & 0.6594 & 0.6631 & 0.6659 & 0.6680 & 0.6697 & 0.6711 & 0.6723 \\ 
       &          &          &  $\Delta$SCF & 0.5672 & 0.5782 &  & 0.6024 & 0.6169 & 0.6252 & 0.6306 & 0.6345 & 0.6374 & 0.6396 & 0.6414 & 0.6429 & 0.6442 \\ 
LiH    & 3.0139 & 0.2850 &  $-\eps_{ho}$  & 0.1614 & 0.1673 & 0.1814 & 0.1939 & 0.2182 & 0.2355 & 0.2485 & 0.2585 & 0.2666 & 0.2732 & 0.2787 & 0.2834 & 0.2874 \\ 
       &          &          &  $-\eps_{e-ho}$ & 0.3377 & 0.3325 & 0.3359 & 0.3394 & 0.3462 & 0.3508 & 0.3541 & 0.3566 & 0.3585 & 0.3601 & 0.3614 & 0.3624 & 0.3633 \\ 
       &          &          &  $\Delta$SCF & 0.2814 & 0.2829 &  & 0.2940 & 0.3032 & 0.3092 & 0.3134 & 0.3166 & 0.3191 & 0.3212 & 0.3229 & 0.3244 & 0.3257 \\ 
Li$_2$ & 5.0518 & 0.1879 &  $-\eps_{ho}$  & 0.1189 & 0.1202 & 0.1246 & 0.1286 & 0.1368 & 0.1430 & 0.1480 & 0.1522 & 0.1557 & 0.1588 & 0.1614 & 0.1638 & 0.1659 \\ 
       &          &          &  $-\eps_{e-ho}$ & 0.2096 & 0.2073 & 0.2076 & 0.2088 & 0.2116 & 0.2137 & 0.2154 & 0.2167 & 0.2178 & 0.2187 & 0.2195 & 0.2201 & 0.2207 \\ 
       &          &          &  $\Delta$SCF & 0.1875 & 0.1855 &  & 0.1884 & 0.1918 & 0.1943 & 0.1961 & 0.1976 & 0.1988 & 0.1998 & 0.2007 & 0.2014 & 0.2021 \\ 
LiF    & 2.9553 & 0.4155 &  $-\eps_{ho}$  & 0.2333 & 0.2449 & 0.2708 & 0.2935 & 0.3363 & 0.3655 & 0.3867 & 0.4029 & 0.4156 & 0.4259 & 0.4344 & 0.4416 & 0.4477 \\ 
       &          &          &  $-\eps_{e-ho}$  & 0.5808 & 0.5636 & 0.5511 & 0.5463 & 0.5423 & 0.5412 & 0.5407 & 0.5405 & 0.5402 & 0.5400 & 0.5398 & 0.5396 & 0.5395 \\ 
       &          &          &  $\Delta$SCF & 0.4518 & 0.4355 &  & 0.4200 & 0.4173 & 0.4170 & 0.4172 & 0.4176 & 0.4180 & 0.4184 & 0.4189 & 0.4193 & 0.4198 \\ 
BeH    & 2.5368 & 0.3015 &  $-\eps_{ho}$  & 0.2441 & 0.2471 & 0.2543 & 0.2603 & 0.2711 & 0.2782 & 0.2833 & 0.2871 & 0.2901 & 0.2925 & 0.2945 & 0.2963 & 0.2977 \\ 
       &          &          &  $-\eps_{e-ho}$ & 0.3263 & 0.3185 & 0.3142 & 0.3132 & 0.3133 & 0.3140 & 0.3148 & 0.3154 & 0.3159 & 0.3164 & 0.3168 & 0.3171 & 0.3174 \\ 
       &          &          &  $\Delta$SCF & 0.3116 & 0.3040 &  & 0.3013 & 0.3027 & 0.3039 & 0.3049 & 0.3056 & 0.3062 & 0.3067 & 0.3071 & 0.3075 & 0.3077 \\ 
BH     & 2.3290 & 0.359  &  $-\eps_{ho}$  & 0.2031 & 0.2121 & 0.2281 & 0.2412 & 0.2654 & 0.2822 & 0.2946 & 0.3043 & 0.3121 & 0.3185 & 0.3239 & 0.3286 & 0.3325 \\ 
       &          &          &  $-\eps_{e-ho}$ & 0.3818 & 0.3802 & 0.3823 & 0.3851 & 0.3910 & 0.3949 & 0.3978 & 0.3999 & 0.4015 & 0.4028 & 0.4039 & 0.4047 & 0.4055 \\ 
       &          &          &  $\Delta$SCF & 0.3477 & 0.3478 &  & 0.3536 & 0.3594 & 0.3632 & 0.3659 & 0.3679 & 0.3693 & 0.3705 & 0.3715 & 0.3723 & 0.3729 \\ 
BO     & 2.2766 & 0.489 &  $-\eps_{ho}$  & 0.3499 & 0.3592 & 0.3786 & 0.3953 & 0.4265 & 0.4482 & 0.4599 & 0.4687 & 0.4756 & 0.4812 & 0.4858 & 0.4897 & 0.4930 \\ 
       &          &          &  $-\eps_{e-ho}$ & 0.5250$^{\dagger}$ & 0.5152$^{\dagger}$ & 0.5110$^{\dagger}$ & 0.5114$^{\dagger}$ & 0.5155$^{\dagger}$ & 0.5193$^{\dagger}$ & 0.5222$^{\ddagger}$ & 0.5245$^{\ddagger}$ & 0.5263$^{\ddagger}$ & 0.5277$^{\ddagger}$ & 0.5289$^{\ddagger}$ & 0.5298$^{\ddagger}$ & 0.5306$^{\ddagger}$ \\ 
       &          &          &  $\Delta$SCF & 0.4827 & 0.4771 &  & 0.4799 & 0.4858 & 0.4902 & 0.4933 & 0.4957 & 0.4976 & 0.4991 & 0.5004 & 0.5014 & 0.5023 \\ 
BF     & 2.3861 & 0.4087 &  $-\eps_{ho}$  & 0.2509 & 0.2605 & 0.2778 & 0.2920 & 0.3183 & 0.3364 & 0.3498 & 0.3601 & 0.3684 & 0.3753 & 0.3810 & 0.3859 & 0.3901 \\ 
       &          &          &  $-\eps_{e-ho}$ & 0.4345 & 0.4336 & 0.4368 & 0.4407 & 0.4483 & 0.4534 & 0.4569 & 0.4595 & 0.4615 & 0.4631 & 0.4644 & 0.4654 & 0.4663 \\ 
       &          &          &  $\Delta$SCF & 0.3953 & 0.3960 &  & 0.4036 & 0.4111 & 0.4162 & 0.4197 & 0.4224 & 0.4244 & 0.4261 & 0.4274 & 0.4285 & 0.4295 \\ 
CH     & 2.1163 & 0.391 &  $-\eps_{ho}$  & 0.2395 & 0.2459 & 0.2620 & 0.2762 & 0.3031 & 0.3218 & 0.3357 & 0.3465 & 0.3551 & 0.3623 & 0.3682 & 0.3734 & 0.3778 \\ 
       &          &          &  $-\eps_{e-ho}$ & 0.4518 & 0.4386 & 0.4332 & 0.4327 & 0.4342 & 0.4360 & 0.4375 & 0.4387 & 0.4396 & 0.4404 & 0.4410 & 0.4416 & 0.4421 \\ 
       &          &          &  $\Delta$SCF & 0.4099 & 0.3978 &  & 0.3916 & 0.3925 & 0.3939 & 0.3952 & 0.3962 & 0.3971 & 0.3979 & 0.3985 & 0.3990 & 0.3995 \\ 
CN     & 2.2144 & 0.4997 &  $-\eps_{ho}$  & 0.3524 & 0.3601 & 0.3774 & 0.3923 & 0.4203 & 0.4398 & 0.4543 & 0.4656 & 0.4746 & 0.4822 & 0.4884 & 0.4938 & 0.4985 \\ 
       &          &          &  $-\eps_{e-ho}$ & 0.5942 & 0.5841 & 0.5786 & 0.5771 & 0.5766 & 0.5770 & 0.5773 & 0.5775 & 0.5776 & 0.5777 & 0.5777 & 0.5777 & 0.5777 \\ 
       &          &          &  $\Delta$SCF & 0.5433 & 0.5416 &  & 0.5521 & 0.5634 & 0.5714 & 0.5773 & 0.5820 & 0.5857 & 0.5887 & 0.5913 & 0.5934 & 0.5953 \\ 
CO     & 2.1322 & 0.515 &  $-\eps_{ho}$  & 0.3350 & 0.3496 & 0.3745 & 0.3946 & 0.4309 & 0.4556 & 0.4737 & 0.4876 & 0.4988 & 0.5079 & 0.5155 & 0.5219 & 0.5275 \\ 
       &          &          &  $-\eps_{e-ho}$ & 0.5717 & 0.5702 & 0.5743 & 0.5794 & 0.5896 & 0.5965 & 0.6014 & 0.6050 & 0.6077 & 0.6099 & 0.6116 & 0.6130 & 0.6142 \\ 
       &          &          &  $\Delta$SCF & 0.5126 & 0.5134 &  & 0.5231 & 0.5326 & 0.5390 & 0.5434 & 0.5467 & 0.5492 & 0.5512 & 0.5529 & 0.5542 & 0.5553 \\ 
NH     & 1.9600 & 0.4958 &  $-\eps_{ho}$  & 0.3157 & 0.3320 & 0.3573 & 0.3770 & 0.4093$^{\ddagger}$ & 0.4305$^{\ddagger}$ & 0.4461$^{\ddagger}$ & 0.4581$^{\ddagger}$ & 0.4678$^{\ddagger}$ & 0.4757$^{\ddagger}$ & 0.4824$^{\ddagger}$ & 0.4880$^{\ddagger}$ & 0.4929$^{\ddagger}$ \\ 
       &          &          &  $-\eps_{e-ho}$ & 0.5586 & 0.5571 & 0.5590 & 0.5618 & 0.5632 & 0.5633 & 0.5636 & 0.5638 & 0.5640 & 0.5641 & 0.5643 & 0.5644 & 0.5644 \\ 
       &          &          &  $\Delta$SCF & 0.4846 & 0.4872 &  & 0.4968 & 0.5047 & 0.5099 & 0.5134 & 0.5161 & 0.5181 & 0.5196 & 0.5209 & 0.5219 & 0.5228 \\ 
N$_2$  & 2.0743 & 0.5733 &  $-\eps_{ho}$  & 0.3825 & 0.3972 & 0.4239 & 0.4456 & 0.4849 & 0.5115 & 0.5311 & 0.5463 & 0.5584 & 0.5684 & 0.5767 & 0.5839 & 0.5901 \\ 
       &          &          &  $-\eps_{e-ho}$ & 0.6252 & 0.6229 & 0.6271 & 0.6327 & 0.6443 & 0.6526 & 0.6586 & 0.6631 & 0.6667 & 0.6696 & 0.6719 & 0.6739 & 0.6755 \\ 
       &          &          &  $\Delta$SCF & 0.5745 & 0.5735 &  & 0.5846 & 0.5964 & 0.6046 & 0.6105 & 0.6149 & 0.6184 & 0.6211 & 0.6234 & 0.6253 & 0.6270 \\ 
NO     & 2.1746 & 0.3405 &  $-\eps_{ho}$  & 0.1758 & 0.1865 & 0.2116 & 0.2331 & 0.2728 & 0.2997 & 0.3194 & 0.3345 & 0.3466 & 0.3564 & 0.3646 & 0.3715 & 0.3775 \\ 
       &          &          &  $-\eps_{e-ho}$ & 0.4184 & 0.4092 & 0.4093 & 0.4132 & 0.4228 & 0.4298 & 0.4349 & 0.4388 & 0.4418 & 0.4442 & 0.4462 & 0.4478 & 0.4492 \\ 
       &          &          &  $\Delta$SCF & 0.3698 & 0.3615 &  & 0.3659 & 0.3752 & 0.3819 & 0.3868 & 0.3905 & 0.3934 & 0.3958 & 0.3977 & 0.3993 & 0.4007 \\ 
OH     & 1.8324 & 0.4784 &  $-\eps_{ho}$  & 0.2747 & 0.2911 & 0.3208 & 0.3448 & 0.3875 & 0.4158 & 0.4362 & 0.4516 & 0.4638 & 0.4736 & 0.4818 & 0.4887 & 0.4946 \\ 
       &          &          &  $-\eps_{e-ho}$ & 0.5826 & 0.5711 & 0.5667 & 0.5669 & 0.5702 & 0.5731 & 0.5752 & 0.5768 & 0.5781 & 0.5790 & 0.5798 & 0.5804 & 0.5809 \\ 
       &          &          &  $\Delta$SCF & 0.4815 & 0.4749 &  & 0.4760 & 0.4814 & 0.4853 & 0.4880 & 0.4900 & 0.4915 & 0.4926 & 0.4935 & 0.4942 & 0.4948 \\ 
O$_2$  & 2.2819 & 0.4531 &  $-\eps_{ho}$  & 0.2641 & 0.2780 & 0.3078 & 0.3329 & 0.3787 & 0.4098 & 0.4324 & 0.4498 & 0.4635 & 0.4748 & 0.4841 & 0.4921 & 0.4989 \\ 
       &          &          &  $-\eps_{e-ho}$ & 0.5300 & 0.5219 & 0.5232 & 0.5281 & 0.5398 & 0.5484 & 0.5547 & 0.5595 & 0.5633 & 0.5663 & 0.5688 & 0.5709 & 0.5727 \\ 
       &          &          &  $\Delta$SCF & 0.4697 & 0.4628 &  & 0.4699 & 0.4816 & 0.4902 & 0.4965 & 0.5014 & 0.5052 & 0.5083 & 0.5109 & 0.5130 & 0.5149 \\ 
FH     & 1.7326 & 0.5924 &  $-\eps_{ho}$  & 0.3608 & 0.3792 & 0.4129 & 0.4403 & 0.4895 & 0.5221 & 0.5455 & 0.5633 & 0.5772 & 0.5885 & 0.5978 & 0.6057 & 0.6124 \\ 
       &          &          &  $-\eps_{e-ho}$ & 0.7363 & 0.7222 & 0.7135 & 0.7108 & 0.7095 & 0.7096 & 0.7098 & 0.7100 & 0.7101 & 0.7101 & 0.7101 & 0.7101 & 0.7101 \\ 
       &          &          &  $\Delta$SCF & 0.6179 & 0.6250 &  & 0.6576 & 0.6845 & 0.7024 & 0.7152 & 0.7248 & 0.7322 & 0.7382 & 0.7430 & 0.7471 & 0.7505 \\ 
F$_2$  & 2.6695 & 0.5769 &  $-\eps_{ho}$  & 0.3544 & 0.3741 & 0.4118 & 0.4425 & 0.4972 & 0.5333 & 0.5591 & 0.5785 & 0.5937 & 0.6060 & 0.6162 & 0.6247 & 0.6320 \\ 
       &          &          &  $-\eps_{e-ho}$ & 0.6473 & 0.6411 & 0.6447 & 0.6513 & 0.6660 & 0.6763 & 0.6839 & 0.6895 & 0.6939 & 0.6975 & 0.7004 & 0.7028 & 0.7048 \\ 
       &          &          &  $\Delta$SCF & 0.5748 & 0.5848 &  & 0.6293 & 0.6660 & 0.6901 & 0.7072 & 0.7200 & 0.7300 & 0.7381 & 0.7447 & 0.7502 & 0.7549 \\ 
H      &          & 0.4997 &  $-\eps_{ho}$  & 0.5000 & 0.5000 & 0.5000 & 0.5000 & 0.5000 & 0.5000 & 0.5000 & 0.5000 & 0.5000 & 0.5000 & 0.5000 & 0.5000 & 0.5000 \\ 
       &          &          &  $-\eps_{e-ho} $& 0.5000 & 0.5000 & 0.5000 & 0.5000 & 0.5000 & 0.5000 & 0.5000 & 0.5000 & 0.5000 & 0.5000 & 0.5000 & 0.5000 & 0.5000 \\ 
       &          &          &  $\Delta$SCF & 0.5000 & 0.5000 &  & 0.5000 & 0.5000 & 0.5000 & 0.5000 & 0.5000 & 0.5000 & 0.5000 & 0.5000 & 0.5000 & 0.5000 \\ 
Li     &          & 0.1981 &  $-\eps_{ho}$   & 0.1788 & 0.1780 & 0.1786 & 0.1797 & 0.1821 & 0.1838 & 0.1852 & 0.1863 & 0.1873 & 0.1881 & 0.1888 & 0.1894 & 0.1900 \\ 
       &          &          &  $-\eps_{e-ho}$ & 0.2058 & 0.1982 & 0.1962 & 0.1963 & 0.1972 & 0.1978 & 0.1983 & 0.1987 & 0.1990 & 0.1993 & 0.1995 & 0.1996 & 0.1997 \\ 
       &          &          &  $\Delta$SCF  & 0.2056 & 0.1979 &  & 0.1962 & 0.1971 & 0.1978 & 0.1983 & 0.1986 & 0.1989 & 0.1991 & 0.1993 & 0.1994 & 0.1996 \\ 
Be     &          & 0.3426 &  $-\eps_{ho}$   & 0.2056 & 0.2118 & 0.2225 & 0.2312 & 0.2473 & 0.2586 & 0.2671 & 0.2739 & 0.2794 & 0.2841 & 0.2881 & 0.2915 & 0.2945 \\ 
       &          &          &  $-\eps_{e-ho}$ & 0.3360 & 0.3351 & 0.3381 & 0.3411 & 0.3466 & 0.3504 & 0.3531 & 0.3551 & 0.3567 & 0.3580 & 0.3591 & 0.3600 & 0.3608 \\ 
       &          &          &  $\Delta$SCF  &  & 0.3227 &  & 0.3297 & 0.3354 & 0.3391 & 0.3417 & 0.3437 & 0.3452 & 0.3465 & 0.3475 & 0.3484 & 0.3491 \\ 
B      &          & 0.3049 &  $-\eps_{ho}$   & 0.1894 & 0.1925 & 0.2015 & 0.2100 & 0.2270 & 0.2395 & 0.2490 & 0.2566 & 0.2628 & 0.2679 & 0.2723 & 0.2761 & 0.2793 \\ 
       &          &          &  $-\eps_{e-ho}$ & 0.3377 & 0.3277 & 0.3236 & 0.3234 & 0.3252 & 0.3270 & 0.3285 & 0.3296 & 0.3305 & 0.3312 & 0.3318 & 0.3323 & 0.3327 \\ 
       &          &          &  $\Delta$SCF  & 0.3213 & 0.3116 &  & 0.3067 & 0.3078 & 0.3093 & 0.3105 & 0.3115 & 0.3122 & 0.3128 & 0.3133 & 0.3137 & 0.3141 \\ 
C      &          & 0.4138 &  $-\eps_{ho}$   & 0.2740 & 0.2802 & 0.2943 & 0.3067 & 0.3302 & 0.3468 & 0.3592 & 0.3688 & 0.3766 & 0.3830 & 0.3884 & 0.3930 & 0.3970 \\ 
       &          &          &  $-\eps_{e-ho}$ & 0.4737 & 0.4617 & 0.4555 & 0.4540 & 0.4536 & 0.4541 & 0.4546 & 0.4549 & 0.4552 & 0.4555 & 0.4557 & 0.4558 & 0.4560 \\ 
       &          &          &  $\Delta$SCF  & 0.4384 & 0.4269 &  & 0.4184 & 0.4175 & 0.4176 & 0.4179 & 0.4181 & 0.4183 & 0.4184 & 0.4185 & 0.4186 & 0.4186 \\ 
N      &          & 0.5341 &  $-\eps_{ho}$   & 0.3685 & 0.3780 & 0.3974 & 0.4138 & 0.4438 & 0.4645 & 0.4796 & 0.4912 & 0.5005 & 0.5082 & 0.5145 & 0.5199 & 0.5246 \\ 
       &          &          &  $-\eps_{e-ho}$ & 0.6158 & 0.6019 & 0.5944 & 0.5923 & 0.5913 & 0.5922 & 0.5928 & 0.5933 & 0.5937 & 0.5940 & 0.5942 & 0.5944 & 0.5946 \\ 
       &          &          &  $\Delta$SCF  & 0.5590 & 0.5448 &  & 0.5356 & 0.5349 & 0.5353 & 0.5358 & 0.5362 & 0.5365 & 0.5367 & 0.5369 & 0.5371 & 0.5372 \\ 
O      &          & 0.5005 &  $-\eps_{ho}$   & 0.2753 & 0.2941 & 0.3272 & 0.3536 & 0.4001 & 0.4308 & 0.4526 & 0.4690 & 0.4819 & 0.4924 & 0.5010 & 0.5083 & 0.5145 \\ 
       &          &          &  $-\eps_{e-ho}$ & 0.5842 & 0.5771 & 0.5769 & 0.5799 & 0.5872 & 0.5924 & 0.5960 & 0.5987 & 0.6007 & 0.6022 & 0.6035 & 0.6046 & 0.6054 \\ 
       &          &          &  $\Delta$SCF  & 0.4947 & 0.4905 &  & 0.4974 & 0.5062 & 0.5121 & 0.5162 & 0.5192 & 0.5214 & 0.5231 & 0.5245 & 0.5257 & 0.5266 \\ 
F      &          & 0.6403 &  $-\eps_{ho}$   & 0.3810 & 0.4025 & 0.4414 & 0.4724 & 0.5269 & 0.5622 & 0.5872 & 0.6060 & 0.6206 & 0.6324 & 0.6420 & 0.6502 & 0.6570 \\ 
       &          &          &  $-\eps_{e-ho}$ & 0.7626 & 0.7491 & 0.7428 & 0.7424 & 0.7449 & 0.7473 & 0.7490 & 0.7502 & 0.7511 & 0.7518 & 0.7523 & 0.7528 & 0.7531 \\ 
       &          &          &  $\Delta$SCF  & 0.6464 & 0.6362 &  & 0.6358 & 0.6415 & 0.6457 & 0.6485 & 0.6505 & 0.6520 & 0.6531 & 0.6539 & 0.6546 & 0.6551 \\  \hline
       
\multicolumn{1}{c}{}&\multicolumn{1}{c}{}&\multicolumn{1}{c}{}&\multicolumn{1}{l}{$\delta_{IP}$ (\%) via $-\eps_{ho}$}& \multicolumn{1}{r}{\textbf{36.50}} & \multicolumn{1}{r}{\textbf{34.16}}& \multicolumn{1}{r}{\textbf{29.56}}& \multicolumn{1}{r}{\textbf{25.74}}& \multicolumn{1}{r}{\textbf{18.89}}& \multicolumn{1}{r}{\textbf{14.40}}& \multicolumn{1}{r}{\textbf{11.31}}& \multicolumn{1}{r}{\textbf{9.14}}& \multicolumn{1}{r}{\textbf{7.67}}& \multicolumn{1}{r}{\textbf{6.73}}& \multicolumn{1}{r}{\textbf{6.21}}& \multicolumn{1}{r}{\textbf{6.03}}& \multicolumn{1}{r}{\textbf{6.07}}\\

\multicolumn{1}{c}{}&\multicolumn{1}{c}{}&\multicolumn{1}{c}{}&\multicolumn{1}{l}{$\delta_{IP}$ (\%) via $-\eps_{e-ho}$}& \multicolumn{1}{r}{\textbf{15.83}} & \multicolumn{1}{r}{\textbf{14.04}}& \multicolumn{1}{r}{\textbf{13.56}}& \multicolumn{1}{r}{\textbf{13.81}}& \multicolumn{1}{r}{\textbf{14.73}}& \multicolumn{1}{r}{\textbf{15.50}}& \multicolumn{1}{r}{\textbf{16.09}}& \multicolumn{1}{r}{\textbf{16.54}}& \multicolumn{1}{r}{\textbf{16.89}}& \multicolumn{1}{r}{\textbf{17.18}}& \multicolumn{1}{r}{\textbf{17.41}}& \multicolumn{1}{r}{\textbf{17.61}}& \multicolumn{1}{r}{\textbf{17.78}}\\

\multicolumn{1}{c}{}&\multicolumn{1}{c}{}&\multicolumn{1}{c}{}&\multicolumn{1}{l}{$\delta_{IP}$ (\%) via $\Delta$SCF}& \multicolumn{1}{r}{\textbf{4.13}} & \multicolumn{1}{r}{\textbf{3.19}}& \multicolumn{1}{r}{\textbf{}}& \multicolumn{1}{r}{\textbf{4.23}}& \multicolumn{1}{r}{\textbf{6.05}}& \multicolumn{1}{r}{\textbf{7.39}}& \multicolumn{1}{r}{\textbf{8.37}}& \multicolumn{1}{r}{\textbf{9.13}}& \multicolumn{1}{r}{\textbf{9.72}}& \multicolumn{1}{r}{\textbf{10.20}}& \multicolumn{1}{r}{\textbf{10.60}}& \multicolumn{1}{r}{\textbf{10.94}}& \multicolumn{1}{r}{\textbf{11.22}}\\

\multicolumn{1}{c}{}&\multicolumn{1}{c}{}&\multicolumn{1}{c}{}&\multicolumn{1}{l}{$S$ via $-\eps_{ho}$}& \multicolumn{1}{r}{\textbf{-0.92}} & \multicolumn{1}{r}{\textbf{-0.92}}& \multicolumn{1}{r}{\textbf{-0.92}}& \multicolumn{1}{r}{\textbf{-0.92}}& \multicolumn{1}{r}{\textbf{-0.92}}& \multicolumn{1}{r}{\textbf{-0.92}}& \multicolumn{1}{r}{\textbf{-0.92}}& \multicolumn{1}{r}{\textbf{-0.85}}& \multicolumn{1}{r}{\textbf{-0.62}}& \multicolumn{1}{r}{\textbf{-0.62}}& \multicolumn{1}{r}{\textbf{-0.15}}& \multicolumn{1}{r}{\textbf{0.00}}& \multicolumn{1}{r}{\textbf{0.08}}\\

\multicolumn{1}{c}{}&\multicolumn{1}{c}{}&\multicolumn{1}{c}{}&\multicolumn{1}{l}{$S$ via $-\eps_{e-ho}$}& \multicolumn{1}{r}{\textbf{0.92}} & \multicolumn{1}{r}{\textbf{0.92}}& \multicolumn{1}{r}{\textbf{0.85}}& \multicolumn{1}{r}{\textbf{0.85}}& \multicolumn{1}{r}{\textbf{0.92}}& \multicolumn{1}{r}{\textbf{0.92}}& \multicolumn{1}{r}{\textbf{1.00}}& \multicolumn{1}{r}{\textbf{1.00}}& \multicolumn{1}{r}{\textbf{1.00}}& \multicolumn{1}{r}{\textbf{1.00}}& \multicolumn{1}{r}{\textbf{1.00}}& \multicolumn{1}{r}{\textbf{1.00}}& \multicolumn{1}{r}{\textbf{1.00}}\\
\end{longtable}\vspace{-0.4cm}
\begin{description}
\item[$^\dagger$]\footnotesize In this case, the states marking the highest occupied eigenvalue before and after the shift are not the same, but they belong to different spin channels.
\item[$^\ddagger$] Here, the states marking the highest occupied eigenvalue are of a different spin channel compared to the results obtained with the parameter $c=0$, i.e., a transition occurs.
\end{description}
\newpage
\end{landscape}

%\bibliography{bibliography_ISO,skrefs} 